\documentclass[acmlarge, imwut, authorversion]{acmart}

\AtBeginDocument{%
  \providecommand\BibTeX{{%
    \normalfont B\kern-0.5em{\scshape i\kern-0.25em b}\kern-0.8em\TeX}}}

\usepackage{amsmath}
\usepackage{balance}       
\usepackage{graphics}      
\usepackage[T1]{fontenc}   
\usepackage{lipsum}
\usepackage{color}
\usepackage{colortbl}
\usepackage{booktabs}
\usepackage{textcomp}
\usepackage{microtype}        
\usepackage{ccicons}          
\usepackage{multirow}  
\usepackage{color,soul}  
\usepackage{subcaption}  
\usepackage{listings}  
\usepackage{todonotes}
\usepackage{multirow}

\usepackage[export]{adjustbox} 

\usepackage{array}
\newcolumntype{P}[1]{>{\centering\arraybackslash}p{#1}}

\definecolor{ao(english)}{rgb}{0.0, 0.5, 0.0}

\newcommand{\edit}[1]{{\textcolor{black}{#1}}}

\def\etal{\emph{et al.\ }}

\let\oldsim\sim 
\renewcommand{\sim}{{\oldsim}}

\setcopyright{acmcopyright}
\acmJournal{IMWUT}
\acmYear{2022} \acmVolume{6} \acmNumber{3} \acmArticle{116} \acmMonth{9} \acmPrice{}\acmDOI{10.1145/3550299}

\begin{document}

\title{Assessing the State of Self-Supervised Human Activity Recognition Using Wearables}

\author{Harish Haresamudram}
\email{hharesamudram3@gatech.edu}
\orcid{0000-0002-0545-6504}
\affiliation{%
	\institution{School of Electrical and Computer Engineering, Georgia Institute of Technology}
	\city{Atlanta, GA}
	\country{USA}
}

\author{Irfan Essa}
\email{irfan@gatech.edu}
\orcid{0000-0002-6236-2969}
\affiliation{
	\institution{School of Interactive Computing, Georgia Institute of Technology}
	\city{Atlanta, GA}
	\country{USA}
}

\author{Thomas Pl\"{o}tz}
\email{thomas.ploetz@gatech.edu}
\orcid{0000-0002-1243-7563}
\affiliation{
	\institution{School of Interactive Computing, Georgia Institute of Technology}
	\city{Atlanta, GA}
	\country{USA}
}

\renewcommand{\shortauthors}{Haresamudram et al.}

\begin{abstract}
    The emergence of self-supervised learning in the field of wearables-based human activity recognition (HAR) has opened up opportunities to tackle the most pressing challenges in the field, namely to exploit unlabeled data to derive reliable recognition systems \edit{for scenarios where only small amounts of labeled training samples can be collected.}
	As such, self-supervision, i.e., the paradigm of `pretrain-then-finetune' has the potential to become a strong alternative to the predominant end-to-end training approaches, let alone \edit{hand-crafted features for} the classic activity recognition chain.
	Recently a number of contributions have been made that introduced self-supervised learning into the field of HAR, including, Multi-task self-supervision, Masked Reconstruction, CPC, and SimCLR, to name but a few.
	With the initial success of these methods, the time has come for a systematic inventory and analysis of the potential self-supervised learning has for the field.
	This paper provides exactly that.
	We assess the progress of self-supervised HAR research by introducing a framework that performs a multi-faceted exploration of model performance.
	\edit{
		We organize the framework into three dimensions, each containing three constituent criteria, such that each dimension captures specific aspects of performance, including the robustness to differing source and target conditions, the influence of dataset characteristics, and the feature space characteristics.
		We utilize this framework to assess \edit{seven} state-of-the-art self-supervised methods for HAR, leading to the 
	}
	formulation of insights into the properties of these techniques and to establish their value towards learning representations for diverse scenarios.

\end{abstract}



\maketitle

\section{Introduction}
Driven by the challenges of deriving effective human activity recognition systems from body-worn sensors (HAR), the recent years have seen a significant increase in interest towards learning unsupervised representations of movement data.
While the collection of \textit{annotated} data remains a challenge \cite{kwon2020imutube}, the ubiquitous nature of the sensors themselves allows for the unobtrusive collection of large amounts of \textit{unlabeled} data, which can subsequently be utilized to learn generic representations. 
For example, users can be given a smartwatch equipped with inertial measurement units (IMUs) in order to perform truly in-the-wild data collection of human behaviors.
This practice of training first on large amounts of unlabeled data, followed by the fine-tuning on smaller quantities of labeled data has shown great promise for activity recognition applications~\cite{saeed2019multi, saeed2020federated, saeed2021sense, haresamudram2020masked, haresamudram2021contrastive, tang2020exploring, tang2021selfhar}. 

These self-supervised approaches 
have the potential to alleviate challenges that come with limited annotations (incl.\ overfitting, and availability of less-complex modeling approaches only), and are therefore early proponents of the `pretrain-then-finetune' paradigm shift.
For example, in healthcare, 
collecting large amounts of data may be difficult, due to challenges with participant recruitment.
This is compounded by the necessity to enlist healthcare professionals for annotation, which may incur high costs \cite{hiremath2020deriving}.
In such scenarios, self-supervised methods can jumpstart the training process by first learning relevant weights on large-scale background movement data. 

Methods such as Multi-task self-supervision \cite{saeed2019multi}, Autoencoders \cite{haresamudram2019role}, Masked reconstruction \cite{haresamudram2020masked}, and Contrastive Predictive Coding (CPC) \cite{haresamudram2021contrastive} have been proposed, which aim towards learning generic representations for human activity recognition.
Pre-training with self-supervision is first performed with unlabeled data, followed by fine-tuning of only a simple Multi-Layer Perceptron-based (MLP) classifier using annotations.

For these methods, model pre-training using unlabeled training data is typically either performed utilizing the same dataset as used for training the target activity recognition system--but without using label information--or with a larger background dataset such as Mobiact \cite{chatzaki2016human}.
In either case, both the source and target data are rather similar by, for example, containing samples recorded from similar on-body positions of the movement sensors, and covering roughly the same activities.
Therefore, this evaluation protocol is limited as it only studies one aspect of model performance, which makes it challenging to realize the conditions of effective performance as well as avenues of improvement for these approaches.
To formulate a more well-rounded understanding of these approaches, and to inventory the current state of the field and its overall potential during this paradigm shift, we need to perform a multi-faceted, rigorous evaluation of these approaches in a variety of conditions and scenarios.

In this work, we conduct a large-scale empirical study (\edit{$\sim$50k} activity recognition runs) that evaluates contemporary self-supervised learning methods in HAR across multiple \emph{dimensions} in an effort to "stress test" these methods, so as to shed light on the state-of-the-field. 
Our goal is to deepen the community's understanding of what these approaches learn, and under which conditions they perform best.
The evaluation is performed across three overarching dimensions:
\begin{description}
	\item [1: Robustness to differing source and target conditions] -- wherein the source and target scenarios 	have different data collection conditions, including sensor positions, activities, and sampling rates. 
	Effective performance in this criterion indicates robustness in performance across varying target conditions.

	\item [2: Influence of dataset characteristics] -- which studies the impact of the source and target  characteristics on the activity recognition performance.
	Characteristics such as the class imbalance and quantity of unlabeled and labeled data detail how the setup of the datasets affects downstream recognition tasks. 

	\item [3: Feature space characteristics] -- post-hoc analysis of representations, such as similarity to supervised learning, linear separability, and implicit dimensionality, examines properties of learned representations.
\end{description}
   
We systematically evaluate self-supervised approaches for HAR \cite{saeed2019multi, haresamudram2019role, haresamudram2020masked, haresamudram2021contrastive, tang2020exploring, qian2022makes} on these criteria by first pre-training on the Capture-24 dataset~\cite{chan2021capture, gershuny2020testing, willetts2018statistical}, which contains \edit{data from a single wrist-worn accelerometer} under free-living conditions for 151 participants over $2,500$ hours. 
A standard MLP classifier~\cite{haresamudram2020masked, haresamudram2021contrastive} is then used as the recognition backend.
Through our study, we obtain insights into the workings of these methods under a variety of conditions and determine, for example, that they are not only effective across differing sensor positions, but also to activities not seen during pre-training.
Furthermore, they are data efficient and robust to source dataset imbalances.

The main contributions of this paper can be summarized as follows:
\begin{itemize}
	\item \edit{We provide an overview of the state-of-the-art in the field of self-supervised learning methods for sensor-based human activity recognition. To do so, we conducted an extensive literature survey,  and--based on this--give a concise overview of relevant techniques.}
		
	\item \edit{We introduce an assessment framework comprising of three dimensions, each of which capture a different, relevant aspect of model performance.}
	
	\item \edit{	Using the new assessment framework, we conduct a large-scale experimental evaluation of seven contemporary self-supervised methods for HAR. We derive insights into when and where we can expect particular self-supervised approaches to be useful.}
\end{itemize}

\vspace*{-0.5em}
\section{Background and Motivation}
\label{sec:related}
Human activity recognition (HAR) is typically a five step process, as outlined in the activity recognition chain \cite{reiss2012introducing} -- data collection, pre-processing, segmentation, feature extraction, and classification. 
Here, we focus on the fourth step, which involves the extraction of relevant features for HAR.
There are three main types of representations (or features) in human activity recognition:
\emph{(i)} statistical features: heuristics such as the mean, variance, entropy etc. \cite{huynh2005analyzing, plotz2011feature}; 
\emph{(ii)} distribution-based features, which comprise the state-of-the-art for feature extraction, involving the inverse of the cumulative distribution function~\cite{hammerla2013preserving}; and 
\emph{(iii)} learned features, consisting of dimensionality reduction techniques, supervised and unsupervised (incl.\ self-supervised) learning approaches \cite{ordonez2016deep, hammerla2015pd, saeed2019multi, haresamudram2019role, tang2021selfhar}.

In this work, we focus on methods that learn feature representations from unlabeled sample data.
We perform an assessment of the progress of self-supervised human activity recognition with the goal to gather insights about the approaches and build an understanding of when they work well and where there are opportunities for improvement.
We first discuss the progress made by unsupervised / self-supervised approaches towards learning effective representations for human activity recognition, followed by a brief discussion of prior works designed towards the analysis of self-supervised methods outside HAR and describe how we utilize a selection of them.

\subsection{Self-Supervised Learning in Human Activity Recognition}
\label{sec:background:self-supervised}

\begin{figure*}[t]
	\centering
	\vspace*{-1em}	
	\includegraphics[width=0.75\textwidth]{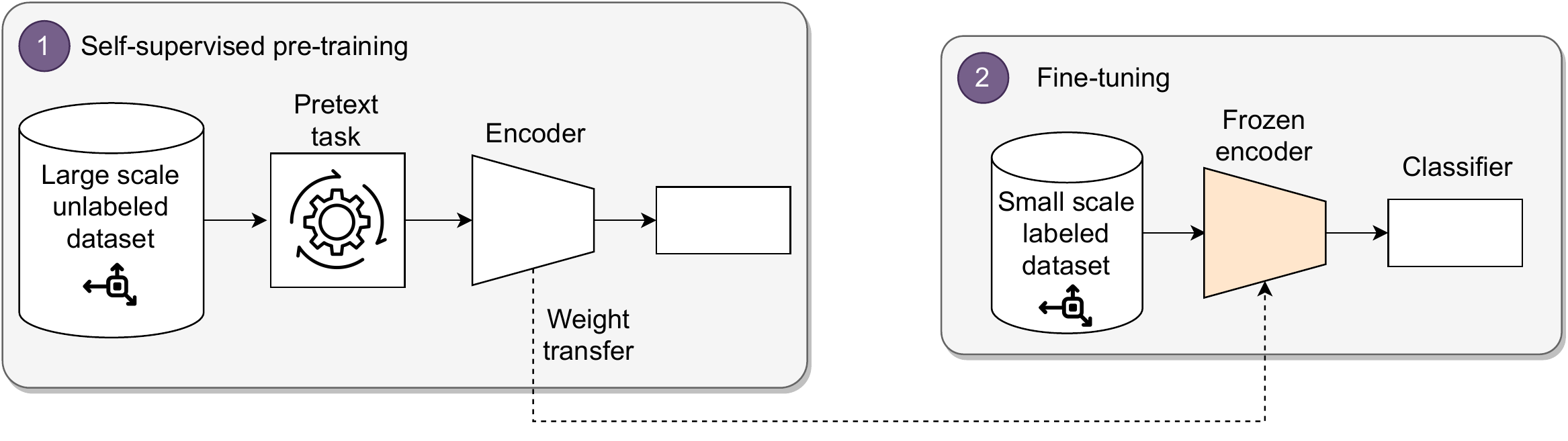}
	\caption{Self-supervised learning comprises a two-stage process where pre-training is first performed with a large-scale unlabeled dataset.
		Subsequently, learned encoder weights are frozen and utilized for HAR on target, small labeled dataset.
	}
	\label{fig:self_sup}
	\vspace*{-1em}
\end{figure*}

Self-supervised learning involves utilizing domain expertise to define `pretext' tasks, which require some semantic understanding. 
They are designed such that they capture relevant characteristics of the input data and by doing so learn weights that are beneficial for downstream recognition.
These methods have shown great promise across many domains, including computer vision \cite{gidaris2018unsupervised}, natural language processing \cite{devlin2018bert, radford2018improving, radford2019language}, and speech processing \cite{chung2019unsupervised, chung2020generative, chung2020improved}. 
They follow the `pretrain-then-finetune' training paradigm (see Fig. \ref{fig:self_sup}) and leverage potentially large quantities of unlabeled data for learning general representations.

\edit{
	A number of such methods have been developed for time-series data in general, typically targeting aspects of time-series properties for learning effective representations.
	In \cite{zerveas2021transformer}, a Transformer encoder \cite{vaswani2017attention} is utilized to reconstruct masked out portions of multi-variate time-series data with the masking being performed across spans of time independently for each channel. 
	Contrastive learning has also been explored for time-series data, by for example, temporally contrasting augmented versions of input \cite{eldele2021time}, by distinguishing signals from within a close neighborhood from ones temporally farther away \cite{tonekaboni2021unsupervised}, or for change point detection \cite{deldari2021time}.
	A wealth of prior work exists in the speech domain, which is closely related to human activity recognition as well due to the time-series and multi-variate nature of the speech representations.
	For example, masked prediction of log-mel spectrograms was studied in \cite{wang2020unsupervised} for speech recognition, with the concept being extended in \cite{zhao2021musicoder, liu2020mockingjay, liu2021tera}.
	Directly predicting data from future timesteps, or autoregressive predictive coding, is a related group of approaches and have proven very effective for speech recognition \cite{chung2020generative, chung2020improved, chung2019unsupervised}.
	Along with them, contrastive learning has also been applied to great success for the same task \cite{schneider2019wav2vec, oord2018representation}. 
}

Self-supervised learning was first explored for human activity recognition in a multi-task setting \cite{saeed2019multi}.
Saeed~\etal~\cite{saeed2019multi} separately applied eight transformations, including adding noise, scaling, rotation, negation, flipping, permutation, time warping and channel-shuffling, with a probability of 50\% in a multi-task setting to accelerometer data from mobile phones.
The primary idea behind such a training scheme is that it allows for capturing core signal characteristics by learning to detect high-level semantics, sensor behavior under different device placements, time-shifting of the events, varying amplitudes, and robustness against sensor noise \cite{saeed2019multi}, thereby aiding in effective representation learning.
The network comprises of a 1D convolutional encoder and task specific fully connected layers, and is trained to independently predict whether each transformation is applied or not. 
After the completion of the pre-training, the encoder layers are frozen and the classifier is evaluated for activity recognition, semi-supervised learning and transfer learning. 
It also contains feature space explorations including the representation similarity, tSNE plots \cite{van2008visualizing}, and saliency maps \cite{simonyan2013deep}

The use of transformer encoders for self-supervision was studied in~\cite{haresamudram2020masked}.
The pretext task involves reconstructing the sensory data only at randomly masked timesteps of windowed accelerometer and gyroscope data from mobile phones.
Ten percent of the timesteps are chosen randomly for masking, and data at those steps is set to zero before being input to the network.
Mean squared error between the sensory data and reconstructed outputs at the masked timesteps is utilized to update network parameters.
This approach seeks to learn useful representations by capturing the local temporal dependencies in the sensory data, by reconstructing data at only the masked timesteps.
Post training, the transformer encoder layers are frozen and only the multi-layer perceptron (MLP) classifier is optimized for activity recognition, and semi-supervised learning.

SelfHAR combined teacher-student self-training and multi-task self-supervision to increase internal and external diversity of training, thereby learning more generalizable features~\cite{tang2021selfhar}. 
Knowledge distillation \cite{hinton2015distilling} is employed to first train the teacher on the labeled dataset.
Subsequently, the learned teacher is used to pseudo-label the large-scale unlabeled dataset, and samples are filtered based on a threshold and top $K$ samples are selected per class.
The combined datasets are used for pre-training with multi-task self-supervision and the learned encoder weights are frozen and used for activity recognition on the original labeled dataset.

In contrast to previous works, SelfHAR also conducts experiments by using the large-scale, wrist-based unlabeled Fenland dataset \cite{lotta2018association, o2015cross} and evaluating on the target datasets, which include those collected at the waist, thereby studying performance across sensor locations.
There is also analysis looking into the effect of the composition of the source unlabeled dataset, where subsets are constructed based on the metabolic equivalents (MET) to obtain subsets with active, inactive, or balanced data. 
All subsets contained nearly the same number of samples and Tang \etal~\cite{tang2021selfhar} find that balanced subsets lead to best downstream recognition performance.

The Contrastive Predictive Coding (CPC) framework was adopted and applied to human activity recognition in \cite{haresamudram2021contrastive}.
Windows of accelerometer and gyroscope data from mobile phones are encoded with a 1D convolutional encoder.
A random timestep was chosen within the window and all encoded data prior to the timestep is summarized into a context vector by a gated recurrent unit (GRU) network \cite{chung2014empirical}.
Subsequently, this context vector was used to predict $k$ future timesteps and the network is optimized using the InfoNCE loss \cite{van2018representation}. 
The core idea behind CPC is that predicting only the next timestep needs the model to understand local variations, but predicting farther into the future is more difficult and requires the model to learn the slowly varying features, or the long term signal.
Capturing such long term dependencies results in strong representation learning.
The frozen encoder and GRU weights are used as feature extractor for activity recognition and semi-supervised learning. 

\edit{
	Exploration into adapting and applying other contrastive learning methods which utilize siamese architectures on augmented views of the input has begun for human activity recognition as well.
	Over the last few years, methods such as SimCLR \cite{chen2020simple}, MoCo \cite{he2020momentum}, SimSiam \cite{chen2021exploring}, NNCLR \cite{dwibedi2021little}, and Barlow Twins \cite{zbontar2021barlow} have shown increasing effectiveness for representation learning and constitute of the state-of-the-art in computer vision.
	These techniques were formulated for computer vision, but have (almost) universally applicable frameworks that allow for easy extension into other domains, including time-series.
	As such, some of these approaches have been studied for sensor data as well, for example, \cite{tang2020exploring} applies the SimCLR framework to activity recognition for healthcare, and \cite{shah2021evaluating} investigates the utilization of both SimCLR and BYOL \cite{grill2020bootstrap} (discussed in detail below) to high-dimensional health signals to assess, for example, sleep, heart, and metabolic conditions. 
	The performance of the aforementioned approaches was also evaluated in \cite{qian2022makes}, which first defines the overall contrastive learning framework, followed by an extensive empirical study of these methods, effective augmentations, and suitable backbone architectures.
}

\edit{
	Therefore, in addition to the previously discussed self-supervised methods for HAR, we also investigate the performance of some approaches from the siamese-based contrastive learning family -- namely SimCLR~\cite{chen2020simple, tang2020exploring}, and SimSiam~\cite{chen2021exploring} for learning activity representations.
	Further, we also study the highly effective Bootstrap Your Own Latent (BYOL) approach, which iteratively bootstraps the outputs of one network to function as the targets for enhanced representation learning \cite{grill2020bootstrap}. 
}

\edit{
	In \cite{tang2020exploring}, the SimCLR \cite{chen2020simple} framework was applied to the Motionsense dataset \cite{malekzadeh2018protecting} for learning unsupervised representations.
	Accelerometer data was used in this study, and encoded through a lightweight 1D convolutional encoder identical to \cite{saeed2019multi}.	
	Each window in the batch is randomly augmented in two different ways and subsequently utilized as the positive pair in the training.
	A projection head consisting of a multi-layer perceptron (MLP) projects the encoded representation into a different space where the NT-Xent (normalized temperature-scaled cross entropy) loss is applied for optimizing network parameters. 
	Augmentations are an important part of the SimCLR framework, and eight time-series transformations were utilized from a collection proposed in \cite{um2017data}.
	For any given positive pair, all other pairs in the batch function as negative samples, thereby eliminating the necessity for memory banks or specialized setups.
	The goal of SimCLR is to maximize the agreement between differently augmented versions of the same input. 
	As demonstrated in \cite{tang2020exploring}, this intuition applies to time-series wearable sensor data as well, resulting in strong representations on Motionsense. 
}

\edit{
	Another related approach consists of SimSiam \cite{chen2021exploring}, which further simplifies the siamese contrastive learning setup by not utilizing negative pairs, a momentum encoder, or large batch sizes.
	As detailed in \cite{qian2022makes}, each window in the batch is randomly augmented twice and passed through a common encoder backbone. 
	A prediction MLP (also called the prediction head) is applied to one side whereas the other side has the stop-gradient operation. 
	The model is trained to maximize the similarity between both sides using a cosine distance based symmetric loss. 
	In this approach, the stop gradient operation is of vital importance and prevents the collapse into constant solutions. 
	For facilitating direct comparison with SimCLR and other methods, we utilize the same simple encoder architecture as described in Multi-task self-supervision \cite{saeed2019multi}.
	SimSiam simplifies the siamese contrastive framework by eliminating aspects of it (e.g., negative pairs), with the intuition that maximizing the similarity between positive pairs is sufficient for learning good representations.
}

\edit{
	Finally, we also evaluate BYOL \cite{grill2020bootstrap}, which utilizes on two neural networks, namely the online and target networks, to interact and learn from one another.
	Using an augmented version of the input, the online network learns to predict the outputs of the target network representation of the same input under a different random transformation.
	The online network comprises of a common backbone that outputs the representation, followed by projection and prediction heads. 
	In contrast, the target network does not contain a prediction head, thereby introducing asymmetry into the architecture. 
	The mean squared-error between the normalized predictions and the target representations is minimized to update the online network.
	The stop-gradient operation prevents the update of the target network, for which a slow-moving average is instead utilized, which prevents representation collapse and results in strong features.
}

Other self-supervised approaches in activity recognition include signal correspondence learning with the wavelet transform in a federated learning framework \cite{saeed2020federated}, utilizing eight auxiliary tasks for learning high-level features \cite{saeed2021sense},
\edit{transformation discrimination for reading activity classification \cite{islam2020self}, and performing constrastive learning across different sensor modalities in multi-sensor setups (ColloSSL) \cite{jain2022collossl}. 
}
In most of the \edit{previously discussed} unsupervised and self-supervised approaches  including Autoencoders, Multi-task self-supervision, Masked reconstruction, SimCLR, SimSiam, BYOL and CPC, the source and target datasets are typically the same for most experiments, albeit the pre-training does not use label information.
In some cases, Mobiact \cite{chatzaki2016human} is used for pre-training instead as it is has the largest number of participants. 
Due to this, source and target conditions such as the activities covered (locomotion-style) and the sensor type (e.g., from mobile phones) and locations (the waist/trousers) are largely identical. 
Such an evaluation protocol only covers one aspect of measuring the model performance -- which is under very similar training and testing conditions. 

\section{Assessment Methodology}
\label{sec:method}
The adoption of self-supervised learning has resulted in remarkable performance improvements in domains such as computer vision, natural language and speech processing. 
Particularly in the case of natural language processing, massive, high-quality datasets have been curated, facilitating the development of increasingly complex model architectures. 
With the introduction of increasingly effective models, the performance on established benchmarks saturated producing only minor improvements on old datasets and tasks.
In order to push the field towards new frontiers,
improved benchmarks were introduced such as, for example, General Language Understanding Evaluation (GLUE)~\cite{wang2018glue} which focuses on natural language understanding (NLU), and Generation, Evaluation, and Metrics (GEM) \cite{gehrmann2021gem} for natural language generation (NLG).
Both bechmarks aggregate \emph{multiple} diverse tasks under a single evaluation framework, thereby targeting many aspects of model performance concurrently.
Interestingly, GEM eschews distilling the complex interplay between data, metrics and model outputs to a single value (like GLUE does) to an interactive result exploration system that presents performance across tasks \cite{gehrmann2021gem}.
Such a system facilitates more nuanced interpretations of the results while rejecting hill climbing prevalent in leaderboards.

In this work, we follow in the steps of GLUE and GEM by designing an evaluation framework specifically for HAR scenarios that comprises of a suite of diverse criteria organized in three dimensions. 
By evaluating on multiple criteria, we assess the generalizability of the methods and discover models that perform well across scenarios. 
Specifically, our framework is similar to GEM wherein we perform diverse evaluations, yet do not attempt to reduce the model performance across tasks into a combined value. 
Rather, the aim is to have nuanced analysis of model performance so as to gauge the progress of the domain and uncover areas of progress.
The dimensions are summarized in Tab. \ref{tab:criteria} and detailed below.

\begin{table}[t]
	\centering
	\vspace*{-0.5em}
	\caption{We define three types of evaluation criteria, each capturing a different part of the self-supervised learning pipeline.}
	\vspace*{-0.5em}
	\small
	\begin{tabular}{|P{.3\textwidth}|P{0.35\textwidth}|P{.25\textwidth}|}
		\hline 
		\cellcolor{lightgray}Robustness to Differing Source and Target Conditions & \cellcolor{lightgray}Influence of Dataset Characteristics & \cellcolor{lightgray}Feature Space Characteristics \\ 
		\hline \hline
		Performance across different sensor positions & Effect of dataset imbalance & Similarity to supervised representations\\ 
		\hline
		Performance for disparate activities & Quantity of unlabeled data required & Linear separability of the representations\\ 
		\hline
		Variations in the sampling rate & Performance under the availability of limited annotated data & Implicit dimensionality of the representations \\ 
		\hline
	\end{tabular}
	\label{tab:criteria}
	\vspace*{-1em}
\end{table}

\vspace*{-0.5em}
\subsection{Robustness to Differing Source and Target Conditions}
This dimension captures the impact of differing source and target conditions such as the sensor locations, targeted activities, and sampling rates on activity recognition, thereby indicating the broad applicability of the methods under diverse settings.
The criteria that form this dimension include: 

\begin{description}
	\item [Performance across different sensor positions] -- this criterion evaluates the applicability of self-super\-vised weights when the source data are collected from a different on-body position than the target domain. 
	Effective performance across differing sensor locations allows for leveraging large scale unlabeled datasets--which are typically collected at the wrist (such as Capture-24 \cite{chan2021capture, gershuny2020testing, willetts2018statistical}, Fenland \cite{lotta2018association, o2015cross}, and Biobank \cite{doherty2017large}), to jumpstart recognition for rarer leg-based activities such as classifying freeze of gait.
	
	\item [Performance for disparate activities] -- source and target domains can contain disparate activities; this criterion evaluates if these self-supervised methods can generalize to unseen activities during fine-tuning.
	
	\item [Variations in the sampling rate] -- target datasets can have different sampling rates compared to the source dataset, on which the pre-training was performed. 
	This criterion examines whether the target sampling rate must match the source, i.e., if downsampling is required or not. 
\end{description}

\vspace*{-0.5em}
\subsection{Influence of Dataset Characteristics}
This dimension studies the impact of the characteristics, such as size of the unlabeled dataset required for pre-training, and the class imbalance on the activity recognition performance.
Capturing such properties sheds light on optimal source and target domain setups for effective recognition.
It comprises of the following criteria:

\begin{description}
	\item [Quantity of unlabeled data required] -- considers  data efficiency of the self-supervised methods. 
	In particular, it looks at how many training users, or alternatively the number of training windows, are required for learning good pre-trained representations.
	Techniques that require smaller training sets are more advantageous as they result in reduced computational costs, a factor that is very important for wearables.
	
	\item [Effect of dataset imbalance] -- many wearable sensing datasets are inherently imbalanced, containing larger quantities of more easily performed activities such as walking or lying down, relative to physically demanding movements such as skipping with a rope.
	We observe such imbalances in free living as well, where sleep might comprise of 7-8 hours per day, but exercising might happen for much shorter durations (such as < 1 hour).
	Given that imbalance is inherent to the datasets, this criterion analyzes the impact of class imbalance in both source and target domains on activity recognition performance.
	
	\item [Performance under the availability of limited annotated data] -- in numerous target scenarios, the data collection and subsequent annotation maybe difficult due to cost or privacy issues. 
	In such cases, approaches that can first pre-train on a different background dataset and subsequently adapt to a target scenario with limited labeling have great practical value.
	The situation is of great importance as it can occur during real-life deployment wherein users can be directed to provide minimal amounts of specific activity data for improving the model performance by adapting to user idiosyncrasies. 
\end{description}

\subsection{Feature Space Characteristics}
This dimension involves the post-hoc analysis of the learned representations and allows us to gain insight of the properties of the features. 
The following criteria are involved in this dimension:
\begin{description}
	\item [Similarity to supervised learning] -- compares the similarity of the self-supervised representations at each layer to an identical network trained end-to-end. 
	This allows us to get an understanding of which approaches mimic the supervised features the best, and also the point in the network where the self-supervised weights diverge from end-to-end training. 

	\item [Linear separability of the representations] -- analyzes whether the self-supervised methods create a feature space where data points are linearly separable. 
	This gives insight into whether the learned features can enable a multitude of downstream tasks, and are not limited to specific ones that relate directly with the pre-training method.
	
	\item [Implicit dimensionality of the representations] -- captures the true capacity and redundancies in the representations learned by self-supervision \cite{gong2019intrinsic}.
	Indicates the number of dimensions that the features can be reduced to, while still containing as much information as possible.
\end{description}

\section{Setup}
We evaluate the performance of our approach on nine representative benchmark datasets.
They have been chosen carefully to cover a wide range of application scenarios, including ranges of sensor positions, activities covered, and sampling rates employed for data recording.
In what follows, we detail the experimental evaluation protocol, which includes descriptions of the: 
\emph{(i)} Datasets used in our study; 
\emph{(ii)} A brief overview of the self-supervised methods we study; 
\emph{(iii)} Implementation details;  and, 
\emph{(iv)} Data pre-processing steps;  
\emph{(v)} Evaluation metrics.

\subsection{Datasets}
In this work, we perform a large-scale empirical study on self-supervised approaches in human activity recognition, specifically focusing on a single accelerometer setup. 
This choice is motivated by the feasibility of only having a single wearable sensor for many scenarios. 
In order to faciliate the transfer of learned weights, we also utilize one accelerometer for the target datasets as well.
A summary of the datasets used in this study is tabulated in Tab.\ \ref{tab:datasets} and discussed below.

\begin{table}[t!]
	\centering
	\vspace*{-1em}
	\caption{
		Overview of the datasets used for our experimental evaluation.
		Capture-24 comprises the source dataset whereas the remaining nine datasets are the target.
		We chose three datasets per location--waist/trousers, wrist, and leg/ankle--covering a wide variety of activities and dataset sizes.
		Note that the sensor location in this table for Daphnet FoG is specified as `leg/ankle' for uniformity; the actual location is on the shank, just above the ankle.
	}
	\small
	\begin{tabular}{P{2.2cm}|P{1.5cm}|c|c|P{8.0cm}}
		Dataset & Location & \# Users & \# Act. & Activities \\ 
		\hline \hline
		Capture-24 \cite{chan2021capture, gershuny2020testing, willetts2018statistical} & Wrist & 151 & - & Free living \\ 
		\hline
		HHAR \cite{stisen2015smart} & Wrist & 9 & 6 & Biking, sitting, going up and down the stairs, standing, and walking \\ 
		\hline
		Myogym \cite{koskimaki2017myogym} & Wrist & 10 & 31 & Seated cable rows, one-arm dumbbell row, wide-grip pulldown behind the neck, bent over barbell row, reverse grip bent-over row, wide-grip front pulldown, bench press, incline dumbbell flyes, incline dumbbell press and flyes, pushups, leverage chest press , close-grip barbell bench press, bar skullcrusher, triceps pushdown, bench dip, overhead triceps extension, tricep dumbbell kickback, spider curl, dumbbell alternate bicep curl, incline hammer curl, concentration curl, cable curl, hammer curl, upright barbell row, side lateral raise, front dumbbell raise, seated dumbbell shoulder press, car drivers, lying rear delt raise, null \\ 
		\hline
		Wetlab \cite{scholl2015wearables} & Wrist & 22 & 9 & Cutting, inverting, peeling, pestling, pipetting, pouring, stirring, transfer, null \\ 
		\hline
		Mobiact \cite{chatzaki2016human} & Waist/ Trousers & 61 & 11 & Standing, walking, jogging, jumping, stairs up, stairs down, stand to sit, sitting on a chair, sit to stand, car step-in, and car step-out\\ 
		\hline
		Motionsense \cite{malekzadeh2018protecting} & Waist/ Trousers & 24 & 6 & Walking, jogging, going up and down the stairs, sitting and standing \\ 
		\hline
		USC-HAD \cite{zhang2012usc} & Waist/ Trousers & 14 & 12 & Walking - forward, left, right, upstairs, and downstairs, running forward, jumping, sitting, standing, sleeping, and riding the elevator up  and down \\ 
		\hline
		Daphnet FoG \cite{bachlin2009wearable} & Leg/ Ankle  & 10 & 3 & No freeze, freeze, null \\ 
		\hline
		MHEALTH \cite{banos2014mhealthdroid} & Leg/ Ankle & 10 & 13 & Standing, sitting, lying down, walking, climbing up the stairs, waist bend forward, frontal elevation of arms, knees bending, cycling, jogging, running, jump front and back\\ 
		\hline
		PAMAP2 \cite{reiss2012introducing} & Leg/ Ankle & 9 & 12 & Lying, sitting, standing, walking, running, cycling, nordic walking, ascending and descending stairs, vaccuum cleaning, ironing, rope jumping \\ 
		\hline
	\end{tabular}
	\label{tab:datasets}
	\vspace*{-1em}
\end{table}

\subsubsection{Capture-24}
Capture-24 \cite{chan2021capture, gershuny2020testing, willetts2018statistical} is a large-scale dataset containing recordings from 151 participants for approximately one day, resulting in a total of around 4,000 hours of recorded data.\footnote{Downloaded from: https://ora.ox.ac.uk/objects/uuid:99d7c092-d865-4a19-b096-cc16440cd001}
The data is collected from a single Axivity AX3 wrist-worn activity tracker and is sampled at 100Hz.
The annotation was performed with Vicon Autograph wearable cameras and Whitehall II sleep diaries, resulting in more than 2,500 hours of coarsely labeled data. 
The annotations are broken into $> 200$ fine-grained activities or 6 broadly defined activities such as sleep, sit-stand, mixed, walking, vehicle, and bicycling \cite{willetts2018statistical}. 
The authors also provide corresponding mapping between the more fine-grained labels and broad mappings for sleep monitoring, activity levels etc.
In our setup, we randomly subset 90\% of the participants for training, whereas the rest are used for validation. 
Thus, the training set comprises of $135$ participants where as the validation set consists of the remaining $16$ participants.
The dataset composition is imbalanced, with around 75\% corresponding to either sleep or sit-stand, whereas bicycling comprises of only 0.78\% of the train dataset windows, as shown in Fig.\ref{fig:capture_24_composition} in the Appendix.

\subsubsection{Target Datasets}
The aim of this study is to assess the self-supervised methods across multiple dimensions, so as to further our understanding of what they learn. 
In order to facilitate a broad evaluation of these approaches, we curate nine target datasets - with the goal of having a diverse collection of activities, number of participants, sensor locations, dataset sizes, and sensors utilized. 
We note that many of these datasets also contain other sensors, such as gyroscopes or indeed accelerometers recording different sensor locations.
Yet, we only consider accelerometer data collected at specific locations as detailed below, given that our focus is towards studying single sensor setups.  

First, we consider target datasets collected at the wrist: 
\emph{(i)} HHAR \cite{stisen2015smart} that covers locomotion-style activities with data collected at the wrist and the waist using smartwatches and smartphones respectively; 
\emph{(ii)} the Myogym dataset \cite{koskimaki2017myogym}, studying short duration, fine-grained gym exercises such as dumbbell curls and rows, collected using the Myo armband; and 
\emph{(iii)} Wetlab \cite{scholl2015wearables}, which comprises of laboratory-based gestures and movements. 

At the waist/trousers, we study two mobile phone based datasets - Mobiact \cite{chatzaki2016human} and Motionsense \cite{malekzadeh2018protecting}, and USC-HAD \cite{zhang2012usc}, which was recorded using the MotionNode platform. 
In each case, the activities studied typically include locomotion-style activities such as walking, running, lying down etc. 
Mobiact also contains short-term transition activities such as stepping in and out of a car, whereas USC-HAD includes more static activities such as riding in the elevator. 

Finally, the leg/ankle datasets include: \emph{(i)} Daphnet \cite{bachlin2009wearable}, where the goal is to identify the freeze of gait symptoms in Parkinson's disease; \emph{(ii)} MHEALTH \cite{banos2014mhealthdroid}, which details a novel framework for the development of mobile health applications, and covers exercises in addition to locomotion activities; and, \emph{(iii)} PAMAP2 \cite{reiss2012introducing} where we consider 12 activities of daily living such as domestic activities and sportive exercises (nordic walking, running etc.)

Apart from being recorded at different sensor locations and covering a diverse set of activities, they also contain a wide spread of participants, with six of the target datasets containing 10-15 participants.
Mobiact is the most diverse with 61 participants, followed by Motionsense at 24. 
From an activity standpoint, many of these datasets cover locomotion activities such as walking, running, sitting etc whereas Myogym contains 31 fine-grained gym exercises.
Lastly, most of the datasets have class imbalance (see Fig.\ \ref{fig:target_composition} for reference) with Wetlab, Mobiact, MHEALTH and Myogym having the most severe class compositions.

\subsection{Self-supervised HAR Methods}
In this work, we assess approaches belonging to the `pretrain-then-finetune' paradigm wherein representations are first learned on a large-scale background dataset. 
Once trained, the models are freely usable for fine-tuning on diverse downstream scenarios, without requiring extra effort.
Given a focus on such methods, we study the performance of \edit{seven} state-of-the-art self-supervised techniques (detailed descriptions are given in Sec.\ \ref{sec:related}, and key HAR references are listed below): 

\begin{description}
	\item [Multi-task Self-supervision \cite{saeed2019multi}:] involves a multi-task setting and identifying whether signal transformations were applied or not. 
	This was the earliest exploration into self-supervised learning for wearables and utilizes transformations targeted for accelerometry.
	It has also shown impressive performance under semi-supervised and transfer learning settings.
	
	\item [Masked reconstruction \cite{haresamudram2020masked}:] learns useful representations by reconstructing only masked out portions of the input windows, thereby leveraging local temporal dependencies.
	It has demonstrated strong recognition when data from both accelerometer and gyroscope are available.
	
	\item [Contrastive Predictive Coding (CPC) \cite{haresamudram2021contrastive}:] adopted and applied the contrastive predictive coding framework to wearable data, and makes use of long term temporal properties for effective representation learning by predicting multiple future timesteps.
	It performs successfully for activity recognition and semi-supervised learning, especially when multiple sensors are available. 
	
	\item [Autoencoder \cite{saeed2019multi, haresamudram2019role}:] consists of reconstructing the entire input window of sensory data, through an encoder-decoder which are mirror images of each other.
	It has a simple setup and therefore typically functions as the baseline for many self-supervised methods.
		
	\item [\edit{SimCLR: \cite{chen2020simple, tang2020exploring}}] \edit{performs  representation learning by contrasting differently augmented views of the same input window where the negative pairs comprise of data from the same match. 
		The framework is simple and flexible while also leading to strong representations for accelerometer data for locomotion activities.
	}

	\item [\edit{SimSiam: \cite{chen2021exploring}}] \edit{details a simple siamese contrastive learning setup, which learns strong features without utilizing negative pairs, momentum encoders, or large batch sizes, by leveraging the stop gradient operation to prevent collapsing to constant solutions.
	}

	\item [\edit{BYOL: \cite{grill2020bootstrap, shah2021evaluating}}] \edit{comprises of two networks- online and target--that work on different augmented versions of the same input window.
		The network is trained to minimize the mean-squared error loss between the online network's prediction and the target representations.
		Only the online network is optimized via the loss and the target network is updated via a slow moving average of the online network.
	}
	
\end{description}
Notably, we do not include SelfHAR \cite{tang2021selfhar} in our study as it employs self-training in conjunction with self-supervision. 
As described in Sec.\ \ref{sec:background:self-supervised}, the teacher model trained on the labeled target dataset is utilized to pseudo-label and filter the source dataset, followed by self-supervised pre-training.
Therefore, each target dataset results in its own pre-trained model, which limits generalizability and thus does not allow for the self-supervised transfer learning setup utilized in our study.

\subsection{Implementation Details}
We implemented all models using the Pytorch framework \cite{paszke2019pytorch}.\edit{\footnote{\edit{Upon acceptance for publication, we aim to publish the source code of the self-supervised approaches under an open-source license.}}} 
Here, we provide the hyperparameter search spaces for all self-supervised approaches and detail relevant information for each method.

As we are pre-training these techniques on a different dataset than studied in the original papers, we perform extensive hyperparameter tuning to obtain the best performance. 
We perform a random search over available combinations for Tab.\ \ref{tab:diff_locs} and \ref{tab:diff_activities}, as it has been shown to be very effective compared to grid searches \cite{bergstra2012random}, while also not consuming extensive computational resources.
For Masked reconstruction, CPC, Autoencoder, \edit{SimCLR, SimSiam, and BYOL} we randomly sample $20$ sets of hyperparameter combinations during pre-training (the parameter spaces of which are detailed below), whereas for Multi-task self-supervision, we perform a grid search over the learning rates and L2 regularizations as the possible combinations are fewer than $20$.
During activity recognition, we randomly sample $50$ hyperparameter combinations across the pre-trained parameters as well as the classification specific parameters such as the classifier learning rate and weight decay for each target dataset. 
For each combination, we perform 5-fold cross validation to obtain the average F1-score, which is more robust to the particular choice of participants in the splits, resulting in a total of $2,250$ classifier runs per method across target datasets.\footnote{50 combinations $\times$ 5 folds $\times$ 9 datasets}
The best pre-training combination for each dataset is once again trained with five random seeds, and these trained models are utilized for five randomized classifier runs (using the same seeds as pre-training), in order to account for the effect of the choice of random seed.
\edit{
	The parameter search for the activity recognition is performed independently for the MLP classifier and the linear evaluation, in order to find the best possible configurations for both.
}
Subsequently, the mean and standard deviation of the F1-scores are reported in Tab.\ \ref{tab:diff_locs} and \ref{tab:diff_activities}.
Unless specified, all subsequent experiments utilize a similar setup of including both five-fold validation as well as five randomized runs, resulting in \edit{around 50k} pre-training + classifier total runs.
The best combination of parameters for each of these methods across all datasets has been tabulated in Tab.\ \ref{tab:multi_params}-\ref{tab:byol_params} in the Appendix, along with the best overall parameter setup in Tab. \ref{tab:overall_params}.

All pre-training and classification runs are trained for 50 epochs with the pre-training utilizing early stopping at 5 epochs. 
Similar to the original papers, the learning rate for classifiers for CPC, Masked reconstruction, and Autoencoder \cite{haresamudram2019role} reduces by a factor of 0.8 every 10 epochs.
\edit{
	This schedule is also applied for the contrastive approaches such as SimCLR, SimSiam, and BYOL.
	By default, the Adam optimizer \cite{kingma2014adam} is utilized with a batch size of $256$, unless specified differently.
}

\edit{
	The architectures, hyperparameters, and implementation details for all self-supervised approaches and supervised baselines have been detailed in Sec. \ref{sec:imp_details} of the Appendix.
}

\subsection{Data Preparation}
We utilize raw accelerometer data from the source and all target datasets.
We perform no filtering or denoising on the datasets, as deep networks have shown powerful capabilities towards learning from raw data itself~\cite{lecun2015deep}.

We downsample the large-scale Capture-24 dataset to 50 Hz as it reduces the computational load for pre-training. 
Additionally, this is also the lowest native sampling rate in the target datasets. 
We reduced the sampling frequency of all target datasets to match 50 Hz.%
\footnote{The Daphnet FoG dataset was originally captured at 64 Hz and is therefore not downsampled as the sampling rates are comparable.} 
The normalization is performed at a per-channel level such that the train split has zero mean and unit variance. 
The means and variances obtained for the train split are utilized for normalizing the validation split. 

The window size is set to 2 seconds to ensure that any randomly picked window can reasonably capture both longer duration activities (such as standing or walking), as well as more short-term activities including gestures and gym exercises. 
For Capture-24, the overlap is set to zero for efficiency reasons because it is a very large dataset, whereas the target datasets have an overlap of $50\%$ as per usual in HAR applications.

We setup five-fold cross validation for each target dataset. 
Each fold consists of 20\% of randomly chosen users comprising the test split, whereas the remaining 80\% of users are once again separated into the train and validation splits at a 80:20 ratio. 
Overall, every participant appears exactly once as part of the test split and pairwise correlations of analysis windows are effectively eliminated through the user-based splits \cite{hammerla2015let}. 
Finally, the means and variances obtained during the normalization of the Capture-24 training set are applied to the target datasets in order to match the dataset statistics.

\subsection{Performance Metric}
We utilize the test set mean F1-score (i.e., macro F1-score) as the main metric to evaluate performance. 
This is motivated by the substantial class imbalance present in the target datasets (see Fig.\ \ref{fig:target_composition} for reference), and thus we require a metric that is resistant to such bias in the class distribution \cite{powers2020evaluation}.
The accuracy and weighted F1-score are not utilized as they are affected by the skewed class distributions, whereas the unweighted mean F1-score, while not ideal, is a reasonable strategy \cite{plotz2021applying}.

The mean F1-score is computed using:
\begin{equation}
	F_m = \frac{2}{|c|}\sum_{c}^{} \frac{prec_{c} \times recall_{c}}{prec_{c} + recall_{c}}
\end{equation}
where $|c|$ is the number of classes ,and $prec_c$ and $recall_c$ are the precision and recall for each class respectively.

\section{Results}
\label{sec:results}
In this section, we systematically explore the performance of the self-supervised approaches as measured using the dimensions presented in Sec.\ \ref{sec:method}.
The goal of our exploration is to perform a well-rounded assessment of these methods to shed light on the avenues of improvement as well as identifying optimal conditions for effective performance across application boundaries.
First, we evaluate the performance of self-supervised approaches when source and target conditions are not identical, followed by the analysis into how source and target dataset properties affect downstream performance.
Finally, we assess the learned representations via an exploration of the feature space, which will further our understanding of representations learned via self-supervision.

\subsection{Robustness to Differing Source and Target Conditions}
For many applications of wearables, the collection of large-scale labeled datasets may be difficult or downright impossible due to privacy or cost reasons. 
In such cases, it may be possible, however, to leverage datasets collected under different conditions (such as sensor locations or targeted activities) by first pre-training to learn useful weights, and subsequently fine-tuning to the specific smaller scale dataset/application. 

This dimension of our evaluation measures the efficacy of such a training process whereby the utility of using a completely disparate dataset/application for pre-training is measured. 
We begin by evaluating the effect of sensor positions on the recognition performance, followed by studying how variation in the activities recorded between the source and target datasets affects activity recognition. 
Lastly, we investigate whether target datasets with higher native sampling rates need to be downsampled to match the pre-training frequency. 
Overall, these criteria shed light on how universal the self-supervised representation learning methods are, as this evaluation stress tests the limits of the methods' transfer capabilities.

\subsubsection{Performance across Different Sensor Positions}
\label{sec:sensor_positions}
As the Capture-24 dataset is collected at the wrist, we evaluate the learned weights at other on-body locations including the waist/trousers (where people often carry their smartphones), and the leg -- where sensors may be placed to target applications such as detecting freeze of gait in Parkinson`s disease or for measuring fitness.
The DeepConvLSTM network \cite{ordonez2016deep} serves as our \edit{primary} supervised baseline.
\edit{
	We also study the performance on simpler architectures, including a MLP-based (\ref{sec:mlp_classifier}), a 1D convolution-based (\ref{sec:conv_classifier}), and LSTM- and GRU-based classifiers (\ref{sec:lstm_classifier}). 
}
All layers of the network are trained using the target dataset labels. 
For the self-supervised methods, we only fine-tune the weights of the \edit{downstream linear and } multi-layer perceptron (MLP) classifier\edit{s} with the target dataset labels whereas the encoder is pre-trained without any annotations. 
The results of this evaluation are detailed in Tab. \ref{tab:diff_locs}.

\edit{
	Studying the supervised baselines, we observe that the simpler Conv. classifier significantly outperforms the more complex DeepConvLSTM baseline on all datasets apart from Mobiact. 
	By and large, the LSTM classifier and the MLP classifier show similar performance, with the GRU classifier performing better than both.
	As the same MLP classifier architecture is also utilized to perform activity recognition with the learned features, we can directly quantify the impact of the representation learning process.
	Considering the best performing self-supervised methods such as SimCLR and Multi-task self-supervision, we see significant improvements over classifying on raw data for a majority of the datasets (e.g., on PAMAP2, SimCLR shows an increase of approx.~10\% F1-score over utilizing raw data).
	Meanwhile, the performance on the waist-based datasets is comparable to using raw data itself.
	The comparable if not better performance of the self-supervised approaches with MLP classification over the MLP-based activity recognition on raw data clearly showcases the necessity of the self-supervised representation learning methods. 
}

\edit{
	Next, we examine the performance of the self-supervised approaches for linear evaluation, i.e., optimizing on a single fully connected layer on the target datasets.
	Across both the waist- and leg-based datasets, we note that the performance is substantially lower relative to all supervised learning baselines. 
	Only in the case of SimCLR do we obtain performance comparable to the supervised approaches (Mobiact is an exception). 
	As the performance is considerably lower than end-to-end training, we also investigate if a more powerful classifier can make better use of the learned representations, thereby resulting in improved activity recognition performance.
}

\edit{
	For the MLP-based classification, we note many self-supervised approaches such as Multi-task self supervision, CPC, SimCLR, SimSiam, and BYOL perform comparably to the supervised DeepConvLSTM, LSTM-, GRU-, and MLP-based classifiers for the waist-based Motionsense dataset.
	In a similar trend, self-supervision shows similar performance to all supervised baselines for USC-HAD.
	Pre-training is less effective for Mobiact, with DeepConvLSTM significantly outperforming the best self-supervised methods, increasing by over 10\% against CPC and 7\% over SimCLR.
	As mentioned previously, we observe the highest improvements over DeepConvLSTM, LSTM, GRU, and MLP classifiers with PAMAP2, whereas the Conv. classifier performs comparably. 
	The overall trend is that transferring weights from the wrist to the waist results in worsened performance (Mobiact and Motionsense, when compared to the best supervised baseline, i.e., the Conv. classifier) or similar performance (USC-HAD).
	In comparison, transferring from the wrist to leg results in comparable performance (Daphnet FoG and PAMAP2, against the Conv. classifier), or gains (MHEALTH).
}
This is can be reasoned by the fact that many activities such as walking, running etc., which are a part of the MHEALTH and PAMAP2 datasets, have tandem motion between the wrist and the leg.
We posit that this results in effective representations via self-supervision.

\edit{
	On the whole, self-supervised methods with the aid of more powerful classifiers (such as the MLP) perform comparably, if not better, to the supervised baselines (DeepConvLSTM, LSTM and MLP classifiers) even though the source location is at the wrist, whereas the target datasets were collected at the waist/leg. 
}
This result is encouraging as it allows practitioners to utilize large-scale movement datasets (which are typically collected at the wrist, e.g., through smartwatches) for fine-tuning on waist- or leg-based applications. 
\edit{
	It is also interesting to note that the convolutional architecture utilized in many approaches in this study, has long range filters (of sizes 24, 16, and 8), and is likely more apt for the target datasets, as they typically comprise of more static activities such as walking, running, sitting, etc. 
	The superior performance demonstrated by all methods utilizing the specific convolutional encoder architecture (e.g., the Conv. classifier, Multi-task self-supervision, SimCLR, SimSiam, and BYOL) indicate its suitability towards classifying such activities of interest.
	Further, a powerful classifier such as the 3-layer MLP (\ref{sec:mlp_classifier}) is essential for obtaining performance comparable to end-to-end training.
}

\begin{table}[t]
	\centering
	\vspace*{-1em}
	\caption{
		The representation learning performance of the self-supervised approaches across differing sensor positions compared against supervised learning. 
		We perform 5-fold cross validation across 5 randomized runs and report the mean and standard deviation.
		\edit{
			Linear evaluation utilizes uses a single fully connected layer as the classifier whereas the MLP classifier has three layers with batch normalization, dropout and ReLU between successive layers (see Sec. \ref{sec:mlp_classifier} for details).
		}
	}
	\edit{
	\begin{tabular}{c|c|c|c|c|c|c}
		\hline
		& \multicolumn{3}{c|}{Waist} & \multicolumn{3}{c}{Leg} \\ 
		\cline{2-7} 
		\multirow{-2}{*}{Method} & Mobiact & Motionsense & USC-HAD & Daphnet  FOG & MHEALTH & PAMAP2 \\ 
		\hline \hline
		\multicolumn{7}{c}{Supervised baselines} \\ 
		\hline 
		DeepConvLSTM & 82.21 $\pm$0.69 & 84.56 $\pm$0.85 & 53.64 $\pm$0.51 & 53.68 $\pm$2.58 & 45.91 $\pm$0.89& 51.22 $\pm$1.91 \\ 
		\hline
		LSTM classifier & 73.02$\pm$0.48 & 86.74$\pm$0.29 & 53.74$\pm$0.99 & 51.24$\pm$1.55 & 44.39$\pm$0.91 & 48.61$\pm$1.82 \\ 
		\hline
		GRU classifier & 76.34$\pm$0.22 & 87.14$\pm$0.89 & 55.23$\pm$1.09 & 53.96$\pm$1.11 & 45.16$\pm$0.6 & 54.21$\pm$1.24 \\ 
		\hline
		Conv. classifier & 78.93$\pm$0.68 & 89.25$\pm$0.5 & 57.9$\pm$0.62 & 53.41$\pm$0.85 & 48.62$\pm$2.3 & 59.76$\pm$1.53 \\ 
		\hline
		MLP classifier & 74.59$\pm$0.52 & 84.45$\pm$0.39 & 55.59$\pm$1.05 & 49.46$\pm$1.16 & 43.33$\pm$0.93 & 50.02$\pm$0.44 \\ 
		\hline
		%
		\multicolumn{7}{c}{Linear evaluation} \\ 
		\hline 
		Multi-task self.\ sup & 55.51$\pm$2.94 & 74.96$\pm$1.37 & 52.32$\pm$2.68 & 48.55$\pm$1.16 & 36.61$\pm$2.9 & 46.9$\pm$1.14 \\ 
		\hline
		Masked Recons. & 42.27$\pm$3.43 & 61.14$\pm$3.45 & 38.49$\pm$1.82 & 47.16$\pm$1.39 & 22.92$\pm$4.62 & 42.32$\pm$1.63 \\ 
		\hline
		CPC & 63.76$\pm$1.62 & 72.89$\pm$2.06 & 49.25$\pm$1.24 & 45.81$\pm$1.83 & 26.28$\pm$2.06 & 45.84$\pm$1.39 \\ 
		\hline
		Autoencoder & 61.01$\pm$0.63 & 55.13$\pm$3.46 & 40.14$\pm$1.34 & 49.2$\pm$0.27 & 26.93$\pm$1.19 & 50.79$\pm$1.09 \\ 
		\hline
		SimCLR & 66.66$\pm$0.7 & 83.93$\pm$1.78 & 50.57$\pm$4.54 & 51.93$\pm$1.12 & 44.71$\pm$2.17 & 50.75$\pm$2.97 \\ 
		\hline
		SimSiam & 58.32$\pm$2.32 & 71.91$\pm$12.27 & 39.01$\pm$9.03 & 45.45$\pm$4.0 & 32.42$\pm$8.9 & 47.85$\pm$2.48 \\ 
		\hline
		BYOL & 56.09$\pm$2.24 & 66.44$\pm$2.76 & 34.07$\pm$4.82 & 46.14$\pm$3.32 & 30.58$\pm$2.7 & 43.89$\pm$3.35 \\ 
		\hline
		%
		%
		\multicolumn{7}{c}{MLP classifier} \\ 
		\hline 
		Multi-task self.\ sup & 69.71 $\pm$2.03 & 83.18 $\pm$1.26 & 56.63 $\pm$1.34 & 54.16 $\pm$1.12 & 48.05 $\pm$1.05& 58.49 $\pm$3.03 \\ 
		\hline
		Masked Recons. & 54.17 $\pm$1.38 & 75.72 $\pm$1.88 & 45.09 $\pm$0.92 & 52.51 $\pm$1.01 & 47.04 $\pm$0.61& 55.12 $\pm$0.96 \\ 
		\hline
		CPC & 72.91 $\pm$0.99 & 84.74 $\pm$1.14 & 51.37 $\pm$2.43 & 51.16 $\pm$1.0 & 45.49 $\pm$1.27& 52.24 $\pm$1.98 \\ 
		\hline
		Autoencoder & 68.69 $\pm$0.56 & 80.7 $\pm$1.66 & 51.32 $\pm$2.16 & 53.05 $\pm$0.85 & 39.2 $\pm$1.58& 56.88 $\pm$2.04 \\ 
		\hline
		SimCLR & 74.89$\pm$1.6 & 85.6$\pm$2.47 & 53.66$\pm$4.12 & 52.46$\pm$1.44 & 50.51$\pm$1.16 & 60.2$\pm$2.32 \\ 
		\hline
		SimSiam & 72.34$\pm$1.68 & 83.4$\pm$1.64 & 53.57$\pm$2.24 & 50.42$\pm$1.66 & 46.04$\pm$1.77 & 59.64$\pm$4.06 \\ 
		\hline
		BYOL & 69.45$\pm$1.24 & 82.18$\pm$1.2 & 51.01$\pm$2.38 & 51.07$\pm$1.71 & 45.14$\pm$2.39 & 55.84$\pm$1.34 \\ 
		\hline
	\end{tabular}
}
	\vspace*{-1em}
	\label{tab:diff_locs}
\end{table}

\subsubsection{Performance for Disparate Activities}
\label{sec:disp_activities}
This criterion evaluates the effect of variations in the activities covered by the source and target datasets on the self-supervised learning performance.
For example, sufficient quantities of labeled data are difficult to obtain for many medical applications of wearable sensing. 
In such cases, \emph{would it be possible to leverage another dataset with completely disjointed activities for pre-training?}

For this analysis, we consider the datasets whose activities are at least partially not contained in the source dataset across the wrist and leg sensor positions.
For example, the Myogym dataset comprises of specific gym exercises under various orientations (seated, bent over, on stomach, on an incline etc.), some of which may not have been performed by the participants of Capture-24.
Similarly, the Wetlab dataset contains laboratory experiments, which are typically not performed in day-to-day living. 

\edit{
	As in Sec. \ref{sec:sensor_positions}, we study the performance of the self-supervised methods both on a simple linear classifier as well as a more sophisticated MLP. 
	Contrary to the previous section, linear evaluation on the best performing pre-training approaches such as Multi-task self-supervision, CPC, and SimCLR, performs comparably/slightly better than the supervised apparoaches for the wrist-based HHAR dataset. 
	This is true in the case of Myogym as well, with SimCLR obtaining a mean F1-score similar to end-to-end training. 
	For the lab activities in Wetlab, we observe a  drop in performance, with a peak difference of $\sim$10\% between DeepConvLSTM and SimCLR.
}

\edit{
	Further, we once again observe the necessity of having a more sophisticated classifier as the multi-layer perceptron drastically improves the activity recognition performance of the self-supervised methods. 
	In the case of HHAR, which likely has the most overlap with Capture-24, Multi-task self-supervision increases from 51\% to 57.5\% with the use of the MLP. 
	Similarly, CPC and SimCLR show improvements from around 56\% to 58\%. 
	The impact of the MLP-based classification is more pronounced on the fine-grained gym activities (e.g., curls, raises, and dips) present in Myogym, with Autoencoder increasing from 11.4\% to 35.5\%, and BYOL improving from 17\% to 39\%.
	A similar trend is observed for the lab-based gestures in Wetlab as well, with ${>}10\%$ increase shown by CPC, Autoencoder, SimSiam, and BYOL.
	However, Wetlab is an exception wherein the supervised baselines such as DeepConvLSTM and the Conv. classifier clearly outperform the best self-supervised methods.
}
	
\edit{
	MHEALTH and PAMAP2 both have activities that can be expected in daily living, albeit they are collected at the leg/ankle, rather than the wrist. 
	In each case, self-supervision results in significant improvements over DeepConvLSTM for PAMAP2 ($\sim 9\%$ over the mean) and MHEALTH ($\sim 4\%$ over the mean). 
}
It is interesting to note that the self-supervised approaches perform comparably to DeepConvLSTM \edit{and the Conv. classifier} on the Daphnet FoG dataset, which contains freeze of gait symptoms for patients with Parkinson's Disease.
The comparable performance of all self-supervised methods on such rarely accessible medical data demonstrates their generalization capacity and their value towards utilization in conditions where labeled data maybe difficult to obtain. 
\edit{
	Further, we observe how having the same target location (i.e., the wrist) can result in increased performance over all supervised baselines, specially in the case of HHAR and Myogym.
	Aided by more sophisticated classifier architectures, the self-supervised methods trained on daily living data from Capture-24 can generalize well to unseen classes including freeze of gait and gym-based exercises.
	Therefore, all subsequent experiments and results in this work utilize the 3-layer MLP as the activity recognizer, given its overall superior performance and ability to capitalize effectively on the representations learned on daily living data.
}

\begin{table}[t]
	\centering
	\vspace*{-1em}
	\caption{
		The representation learning performance of the self-supervised approaches across different source and target activities, compared against supervised learning. 
		We perform 5-fold cross validation across 5 randomized runs and report the mean and standard deviation.
		\edit{
			Linear evaluation utilizes uses a single fully connected layer as the classifier whereas the MLP classifier has three layers with batch normalization, dropout and ReLU between successive layers (see Sec. \ref{sec:mlp_classifier} for details).
		}
	}
	
	\edit{
	\begin{tabular}{c|ccc|ccc}
		\hline
		& \multicolumn{3}{c|}{Wrist} & \multicolumn{3}{c}{Leg} \\ 
		\cline{2-7} 
		\multirow{-2}{*}{Method} & HHAR & Myogym & Wetlab & Daphnet  FOG & MHEALTH & PAMAP2 \\ 
		\hline \hline
		\multicolumn{7}{c}{Supervised baselines} \\ 
		\hline 
		DeepConvLSTM & 54.39 $\pm$2.28 & 39.9 $\pm$1.05 & 31.0 $\pm$0.68 & 53.68 $\pm$2.58 & 45.91 $\pm$0.89& 51.22 $\pm$1.91 \\ 
		\hline
		LSTM classifier & 37.42$\pm$5.04 & 30.4$\pm$2.75 & 26.95$\pm$0.61 & 51.24$\pm$1.55 & 44.39$\pm$0.91 & 48.61$\pm$1.82 \\ 
		\hline
		GRU classifier & 46.02$\pm$2.1 & 37.99$\pm$1.15 & 30.45$\pm$0.42 & 53.96$\pm$1.11 & 45.16$\pm$0.6 & 54.21$\pm$1.24 \\ 
		\hline
		Conv. classifier & 55.43$\pm$1.21 & 38.63$\pm$0.47 & 30.06$\pm$1.02 & 53.41$\pm$0.85 & 48.62$\pm$2.3 & 59.76$\pm$1.53 \\ 
		\hline
		MLP. classifier & 53.1$\pm$0.81 & 41.06$\pm$0.32 & 24.01$\pm$0.27 & 49.46$\pm$1.16 & 43.33$\pm$0.93 & 50.02$\pm$0.44 \\ 
		\hline
		%
		%
		\multicolumn{7}{c}{Linear evaluation} \\ 
		\hline 
		Multi-task self.\ sup & 50.95$\pm$2.7 & 35.51$\pm$0.38 & 17.39$\pm$1.17 & 48.55$\pm$1.16 & 36.61$\pm$2.9 & 46.9$\pm$1.14 \\ 
		\hline
		Masked Recons. & 43.48$\pm$2.84 & 13.2$\pm$1.26 & 16.63$\pm$0.61 & 47.16$\pm$1.39 & 22.92$\pm$4.62 & 42.32$\pm$1.63 \\ 
		\hline
		CPC & 56.24$\pm$0.98 & 28.54$\pm$2.22 & 14.52$\pm$1.37 & 45.81$\pm$1.83 & 26.28$\pm$2.06 & 45.84$\pm$1.39 \\ 
		\hline
		Autoencoder & 53.57$\pm$1.14 & 11.42$\pm$1.63 & 11.94$\pm$0.59 & 49.2$\pm$0.27 & 26.93$\pm$1.19 & 50.79$\pm$1.09 \\ 
		\hline
		SimCLR & 55.93$\pm$1.75 & 37.9$\pm$2.19 & 20.29$\pm$2.99 & 51.93$\pm$1.12 & 44.71$\pm$2.17 & 50.75$\pm$2.97 \\ 
		\hline
		SimSiam & 45.36$\pm$4.98 & 30.55$\pm$3.79 & 12.76$\pm$3.01 & 45.45$\pm$4.0 & 32.42$\pm$8.9 & 47.85$\pm$2.48 \\ 
		\hline
		BYOL & 40.66$\pm$4.08 & 24.92$\pm$0.64 & 9.9$\pm$0.65 & 46.14$\pm$3.32 & 30.58$\pm$2.7 & 43.89$\pm$3.35 \\ 
		\hline
		%
		%
		\multicolumn{7}{c}{MLP classifier} \\ 
		\hline 
		Multi-task self.\ sup & 57.51 $\pm$1.9 & 42.31 $\pm$2.37 & 23.35 $\pm$0.66 & 54.16 $\pm$1.12 & 48.05 $\pm$1.05& 58.49 $\pm$3.03 \\ 
		\hline
		Masked Recons. & 55.04 $\pm$2.58 & 25.29 $\pm$0.68 & 21.23 $\pm$0.31 & 52.51 $\pm$1.01 & 47.04 $\pm$0.61& 55.12 $\pm$0.96 \\ 
		\hline
		CPC & 58.1 $\pm$1.06 & 39.89 $\pm$0.98 & 24.16 $\pm$0.48 & 51.16 $\pm$1.0 & 45.49 $\pm$1.27& 52.24 $\pm$1.98 \\ 
		\hline
		Autoencoder & 54.25 $\pm$2.04 & 35.45 $\pm$0.49 & 25.75 $\pm$1.03 & 53.05 $\pm$0.85 & 39.2 $\pm$1.58& 56.88 $\pm$2.04 \\ 
		\hline
		SimCLR & 58.55$\pm$2.25 & 42.44$\pm$2.36 & 25.93$\pm$2.24 &  52.46$\pm$1.44 & 50.51$\pm$1.16 & 60.2$\pm$2.32 \\ 
		\hline
		SimSiam & 54.72$\pm$1.32 & 37.63$\pm$3.79 & 23.94$\pm$0.85 & 50.42$\pm$1.66 & 46.04$\pm$1.77 & 59.64$\pm$4.06 \\ 
		\hline
		BYOL & 51.68$\pm$2.27 & 36.48$\pm$1.73 & 20.36$\pm$2.04 & 51.07$\pm$1.71 & 45.14$\pm$2.39 & 55.84$\pm$1.34 \\ 
		\hline
	\end{tabular}
	}
	\label{tab:diff_activities}
	\vspace*{-1em}
\end{table}

\subsubsection{Variations in the Sampling Rate}
The sampling rate of the sensors must be taken in consideration for real-world deployment of wearable systems. 
A low sampling rate conserves computional resources and energy, but trades off against missing relevant signal details that may be vital to discriminating between activities \cite{khan2016optimising}.

\begin{figure}[!t]
	\centering
	\includegraphics[width=0.8\textwidth]{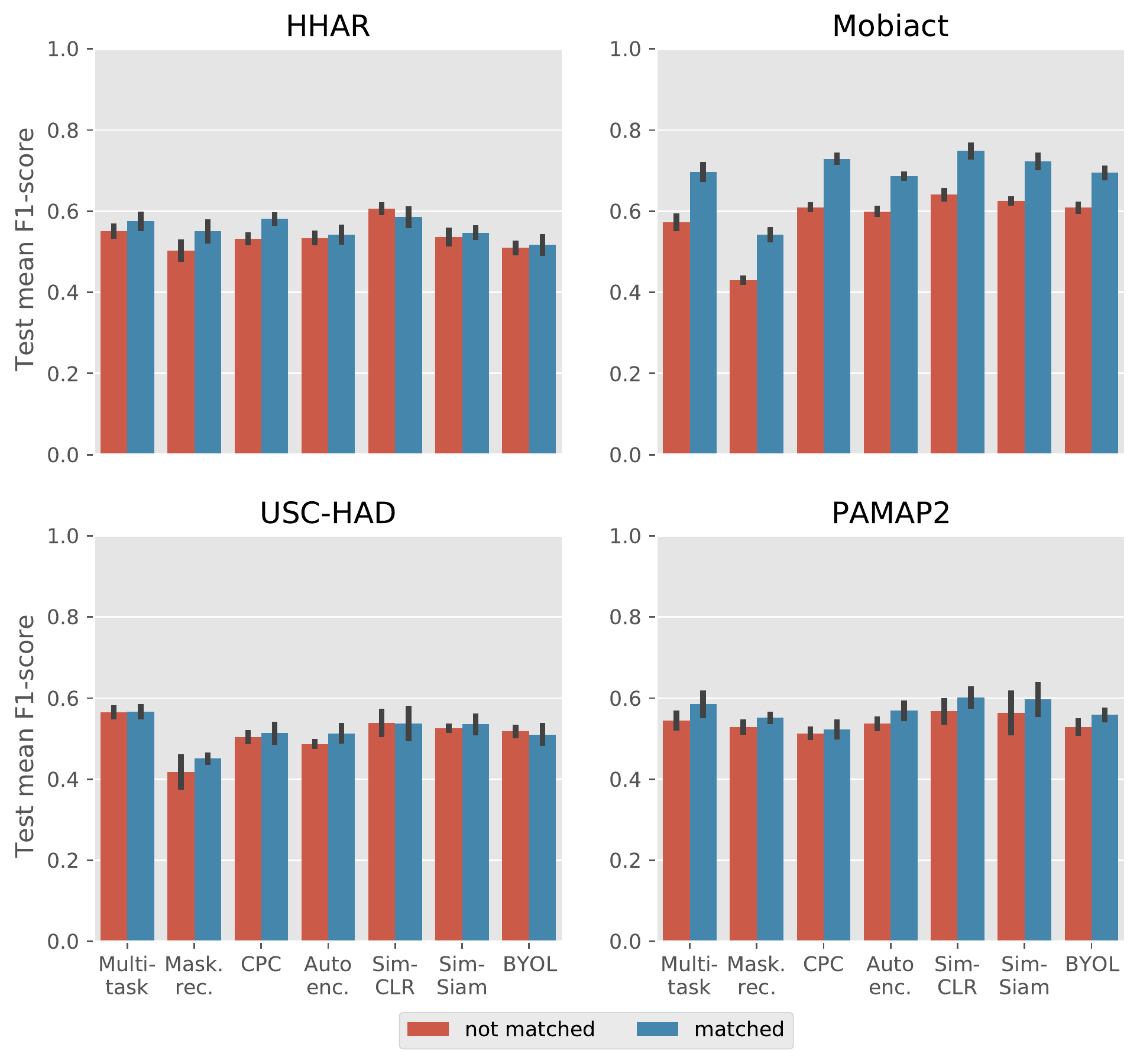}
	\caption{
		Studying the effect of mismatches in  sampling rates of the target datasets to the source dataset: we utilize the best performing models as detailed in Tab. \ref{tab:diff_locs} and \ref{tab:diff_activities} and choose target datasets whose native sampling rate is greater than Capture-24's recording sampling rate. 
		We perform five randomized runs include the pre-training and classification.
		The mean values of the target test set F1-scores are plotted along with the standard deviation for the five runs.
		For the chosen target datasets, the native sampling rate is either 100 or 200 Hz, thereby utilizing half or a quarter of the original pre-training window duration (two seconds).	
		We observe a considerable drop in performance, especially for the Mobiact dataset (200 Hz). 
		As the evaluation window duration is $1/4$ times the pre-training, we observe a considerable drop in performance. 
	}
	\label{fig:downsampled_target}
	\vspace*{-1em}
\end{figure}

In our study, the target datasets are often recorded at higher rates, e.g., the Mobiact dataset was recorded at 200 Hz whereas PAMAP2 was collected at 100 Hz, whereas Capture-24 was downsampled to 50 Hz.
Therefore, we determine if target datasets must be downsampled for matching the sampling rates, and if it is detrimental if the target datasets' native sampling rates are utilized.
Effective performance even when downsampled results in the conservation of the scarce resources available during deployment.
As such, \emph{what is the effect of not matching the target frequency to the source dataset frequency?}
After the pre-training is complete, we compare the activity recognition performance on the target datasets with and without downsampling (shown in Fig. \ref{fig:downsampled_target}).

We observe that all self-supervised methods perform poorly on Mobiact when the sampling rate is not matched to 50 Hz. 
As the original sampling rate of Mobiact is 200Hz, a window size of 100 samples is $0.5$ seconds of data, whereas the pre-training consisted of two seconds of data, thereby resulting in degraded performance. 
The other datasets were recorded at 100 Hz and the reduction in performance occurs to a smaller extent for some of the self-supervised methods. 
Masked reconstruction struggles the most, likely because it depends on surrounding context to predict the missing values. 
If the target dataset is sampled at a different rate, the context looks different than what is seen during pre-training.
Surprisingly, Multi-task self-supervision\edit{, SimCLR, SimSiam, and BYOL are} affected to a significant extent (see Mobiact and PAMAP2) despite having access to a time warping augmentation during pre-training. 
Overall, it is preferable to match the source dataset sampling rates as it different sampling rates result in reduced performance.
The degradation in the F1-score exacerbates with higher frequencies. 

\subsection{Influence of Dataset Characteristics}
The previous dimension evaluated the quality of representations when the target dataset conditions differed from the pre-training setup.
We established that pre-training on a large-scale dataset collected at the wrist has a positive impact on the target activity recognition, even when the sensor locations are different or if the activities are not covered during pre-training. 
A natural extension of the first dimension is to analyze how properties of the datasets, e.g., the number of participants and data collected, affect the downstream recognition performance. 
First, we study the quantity of unlabeled data required for learning effective representations. 
Subsequently, we look at how the class distribution of the source and target datasets impacts the representations learned via self-supervision. 
Finally, we study the performance of these methods when there is limited labeled target data available for fine-tuning. 
Put together, these criteria illuminate the ideal compositions of the datasets, and if extra care must be taken during unlabeled data collection in order to produce the most useful representations possible.

\subsubsection{Quantity of the Unlabeled Data Required}
This criterion evaluates the data efficiency of the self-supervised methods and determines how many participants and what quantities of data overall are necessary for robust learning.
Methods, which are effective without requiring many users or hours of data are preferable \cite{plotz2018deep}.
 
We study the relation between the amount of unlabeled data and the downstream recognition performance from two perspectives:
 \emph{(i)} the number of training participants required; and,
 \emph{(ii)} the number of training windows (across all training users) necessary for effective pre-training. 

\begin{figure}[t]
	\centering
	\vspace*{-1em}
	\includegraphics[width=1\textwidth]{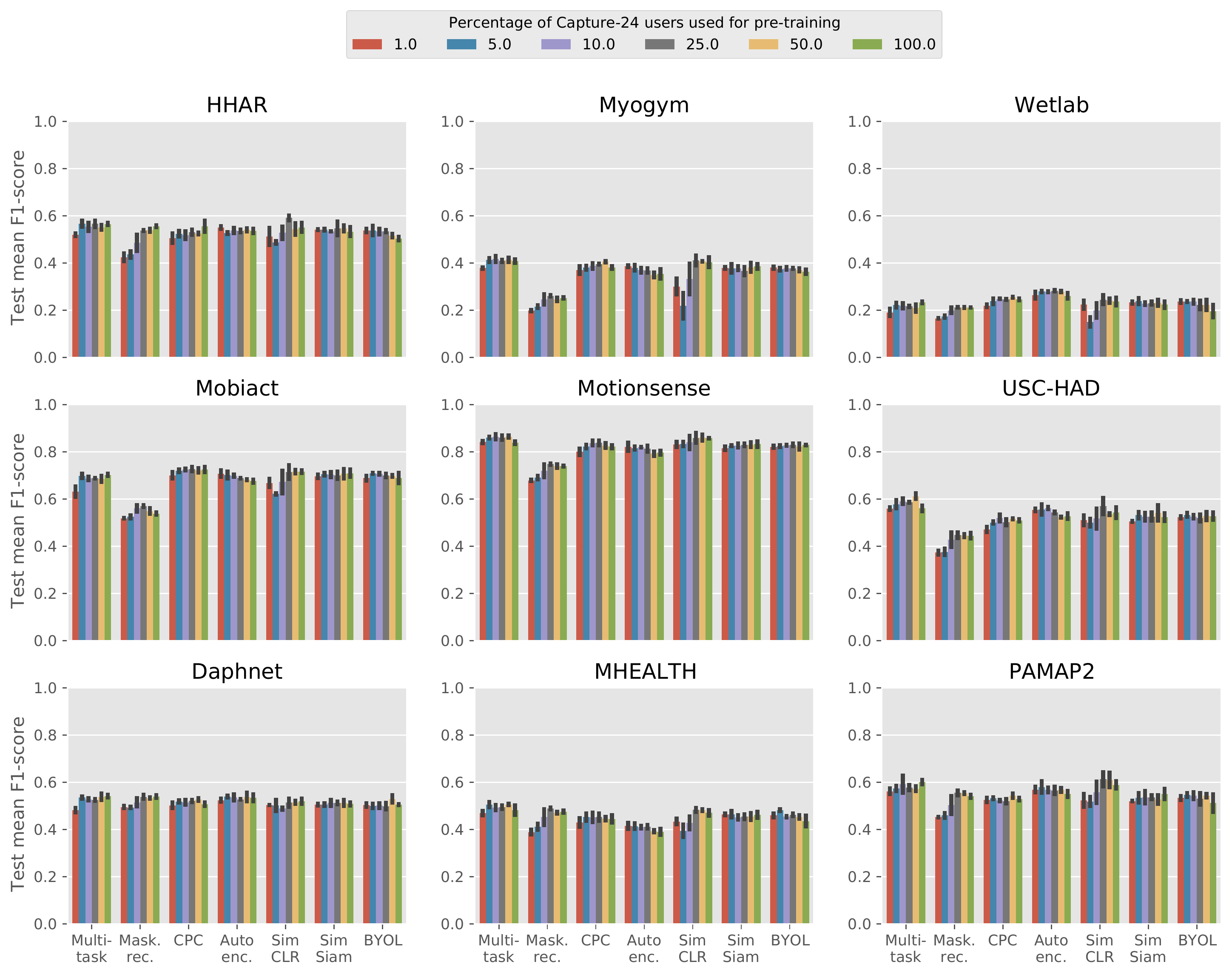}
	\caption{
		Studying the effect of the number of training users on the downstream performance: we utilize the best overall hyperparameter combination (see Tab.\ \ref{tab:overall_params}) and pre-train with the respective approaches while using a varying number of training participants. 
		The validation users for Capture-24 are untouched.
		We perform three randomized runs from choosing the training participants to pre-training, and finally activity recogntition.
		The mean of the target test set F1-scores is plotted along with the standard deviation for three random runs.
		We observe that in most cases, utilizing fewer users results in comparable if not better performance to using the entirety of the training set (which is 135 users).
		Surprisingly, using data from just one user does not result in poor performance; rather, the performance is close to utilizing all available training data.
	}
	\label{fig:perc_users}
	\vspace*{-1em}
\end{figure}

\paragraph{Effect of the number of training users on activity recognition}
Fig.\ \ref{fig:perc_users} details the recognition performance relative to the number of training subjects used for pre-training. 
Out of the $135$ training subjects of Capture-24, we randomly sample $\{1, 5, 10, 25, 50, 100\}\%$ of the participants for pre-training, which corresponds to $\{1, 6, 13, 33, 67, 135\}$ individuals. 
We perform three randomized runs, including choosing training subject(s), pre-training, and classification on the target datasets. We plot the mean and standard deviation of the test set mean F1-score obtained after five-fold validation. 

For all target datasets, we observe that utilizing only one subject for pre-training results in poorer performance. 
This is especially true for Masked Reconstruction, which utilizes a large transformer encoder and thus requires more data.
Adding more subjects improves performance until 50\% of the participants are used for pre-training. 
\edit{
	Specifically for the wrist-based datasets such as HHAR, Myogym, and Wetlab, SimCLR is affected by the lack of diversity in participants, showing an acute drop in performance when only 1 or 6 participants are used for pre-training. 
	In contrast, both SimSiam and BYOL show little change when more participants are available for learning unsupervised representations.
}
Multi-task self-supervision and CPC benefit from larger number of participants but there is generally no significant gain between using 25\% of the subjects over utilizing the entire training set. 
Contrary to \edit{some of} the other approaches, the Autoencoder does not improve with the addition of further participants. 
For example, in the case of datasets such as MHEALTH, Myogym, USC-HAD and Mobiact, the best performance is obtained when only one random participant is for pre-training. 
Thus, \edit{a majority of} the self-supervised methods \edit{including} Multi-task self-supervision, CPC, Autoencoder, \edit{SimCLR, SimSiam, and BYOL} do not require the entire training set (135 participants) for best performance, and can be trained for increased efficiency with fewer participants without much loss in performance.

\paragraph{Effect of the number of training windows on activity recognition}
For data collection, it may not be possible to recruit a large set of participants who are willing to wear  sensor(s) for long periods of time. 
It may be more practical to record data for shorter periods of time in free-living or indeed lab conditions.
In this experiment, we study the feasibility of collecting  a smaller yet diverse (participant-wise) unlabeled dataset with a goal of pre-training for activity recognition. 

\begin{figure}
	\centering
	\vspace*{-1em}
	\includegraphics[width=\textwidth]{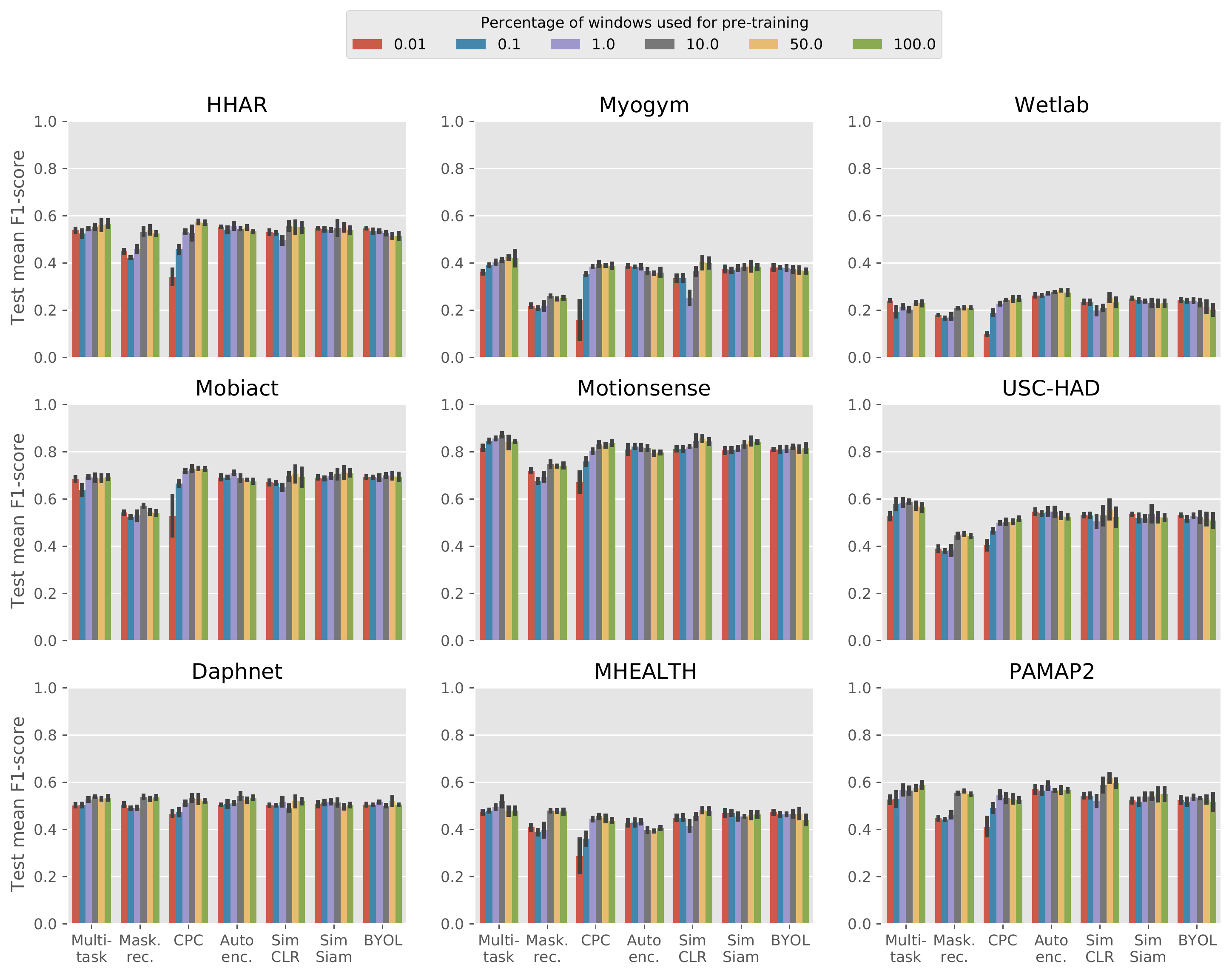}
	\caption{Studying the effect of the number of training windows on the downstream performance: we utilize the best overall hyperparameter combination (see Tab.\ \ref{tab:overall_params}), and pre-train with the respective approaches while using a varying number of training windows. 
	The training windows are sampled randomly from \textbf{all} training participants, while the validation data for Capture-24 are untouched.
	We perform three randomized runs for choosing the training windows, performing pre-training and activity recogntition, and plot the mean of the target test set F1-scores along with the standard deviation.
	We note that excellent performance can be obtained while using 10\% of the training windows, and in some cases using as few as 1\% of the training windows.
	Using a smaller training is advantageous as it can allow for reduced computation costs and for more comprehensive hyperparameter searching.
	}
	\label{fig:perc_windows}
	\vspace*{-1em}
\end{figure}

The Capture-24 dataset contains $135$ participants in the training set, resulting in around 4.1 million non-overlapping windows. 
We create smaller subsets of training data by randomly sub-sampling windows in increasing orders of magnitude of percentages consisting of \{0.01, 0.1, 1, 10, 50, 100\}\% of the training windows from \textbf{all} participants and pre-train self-supervised models, whereas the validation set of Capture-24 remains untouched.
Three randomized runs of pre-training (incl. the randomized subsampling) and classification are performed, and the mean and standard deviation of five-fold test set F1-scores are plotted in Fig. \ref{fig:perc_windows}.
\edit{
	Here, we note that the random sub-sampling of the train data windows preserves the class composition of the full Capture-24 train set. 
	In Fig. \ref{fig:perc_distribution} of the Appendix, we randomly sub-sample \{0.01, 0.1, 1, 10, 50, 100\}\% of the training windows and plot the class distribution.
	For example, the distribution obtained when 0.1\% of the windows are sampled is similar to the class composition of the entire train data. 
	Thus, random sampling does not typically result in picking only one or two classes, which could hamper the self-supervised pre-training specially when those classes may not contain sufficient movement data, e.g., while sleeping or sitting.
}

As can be expected, utilizing only 0.01\% (i.e., $\sim$410) of the training windows results in significantly worse activity recognition. 
This is especially clear for CPC, which shows a near monotonic increase with the addition of more training windows. 
The Multi-task method also performs similarly, getting higher F1-scores with increasing quantities of training windows.
Masked Reconstruction leads to a  jump in performance between using 1\% and 10\% of the windows while the F1-score steadies beyond that.
\edit{
	We note a similar trend for SimCLR as well, with substantial increases in performance between 1\% and 10\%.
	For SimSiam and BYOL, providing more pre-training windows slightly improves the F1-score for most target datasets, without showing drastic changes in performance.
	For Wetlab and MHEALTH, both of these contrastive methods peak in their effectiveness while using only $0.01\%$ of the training windows.
}
\edit{Similarly,} it is interesting to see that the Autoencoder shows the best performance using just $\sim410$ windows on some datasets (such as Myogym, USC-HAD, Mobiact) etc. 
\edit{These methods are} able to learn effective representations with such limited data, showing a similar trend to Fig. \ref{fig:perc_users} wherein the best results were obtained while training one user. 
As a general trend, we observe that the self-supervised methods obtain comparable if not better performance when utilizing only 10\% (or $\sim 400k$) of training windows.

\subsubsection{Effect of the Dataset Imbalance on Performance}
Wearables-based movement datasets tend to be imbalanced, often containing considerably more samples of activities that either occur more naturally or can be performed easily, relative to more physically strenous or rarely occurring activites.
For example, datasets recorded in free-living conditions might contain 7-8 hours of sleep data but contain substantially shorter durations for exercise or commutes. 
\edit{
	Due to this, we study the impact of the source dataset imbalance on downstream HAR.
}

\paragraph{Source dataset imbalance}
\label{sec:source_data_imb}
\begin{figure}
	\centering
	\includegraphics[width=0.8\textwidth]{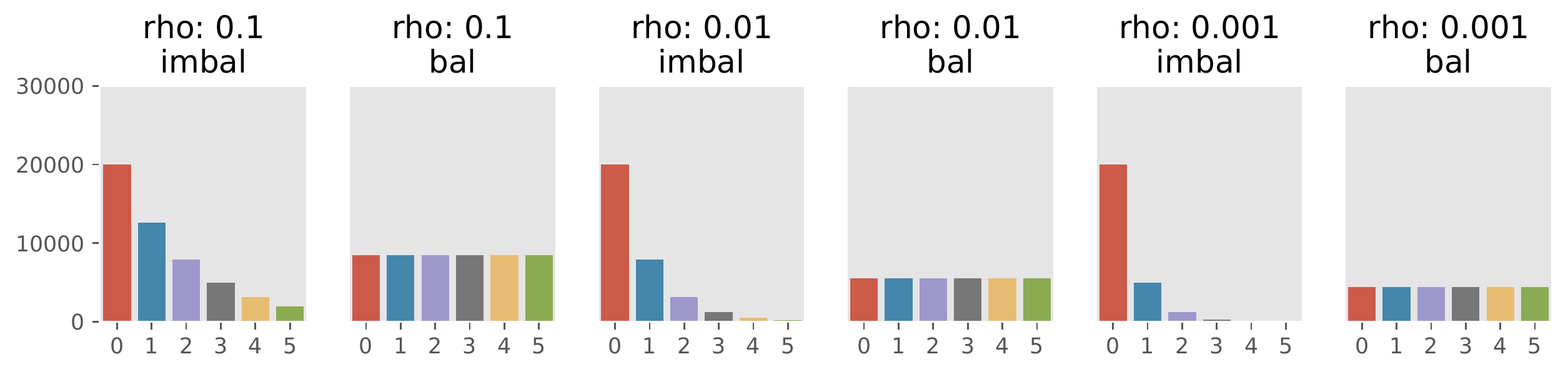}
	\caption{
		Visualizing the class distributions for different values of the imbalance ratio $\rho$.
		We always set the more frequent class to 20000 windows whereas the rarest class is set based on the imbalance ratio.
		The rest of the classes are populated based on the exponential $20000 \times e^{\beta(c-1)}$.
	}
	\label{fig:imb_samples}
	\vspace*{-1em}
\end{figure}

\edit{We adopt the protocol from} \cite{yang2020rethinking, liu2021self} \edit{and} define the imbalance ratio $\rho$ as the ratio of the rarest class to the most frequent class.
We artificially create balanced and imbalanced subsets by using an exponential distribution given by $20,000 \times e^{\beta(c-1)}$ where $20,000$ corresponds to the number of windows of the majority (the most frequent) class, $e$ is the exponential function, $c$ is the class index, and $\beta$ is a scaling factor. 

We study three class imbalance ratios $\rho \in \{0.001, 0.01, 0.1\}$ and obtain $\{\sim27k, \sim33k, \sim51k\}$ training windows. 
\edit{
	The most frequent class is chosen to have 20,000 windows as the resulting imbalanced subset contains $\sim$40k windows, or 1\% of the Capture-24 train set (which was shown to have comparable performance to using the entire train split in Fig. \ref{fig:perc_windows}). 
	More details regarding the setup are present in the Appendix (\ref{app:source_data_imb}).
}

\edit{
	For every $\rho$, the size of the balanced subset matches the imbalanced counterpart, albeit all classes are populated equally.
}
A visualization of this setup is shown in Fig.\ \ref{fig:imb_samples}.
In practice, we perform five randomized runs for the pre-training and in each case, the most frequent and the rarest classes are chosen randomly, thereby making the aggregated performance resistant the choice of the majority class.

\begin{figure}
	\centering
	\vspace*{-1em}
	\includegraphics[width=\textwidth]{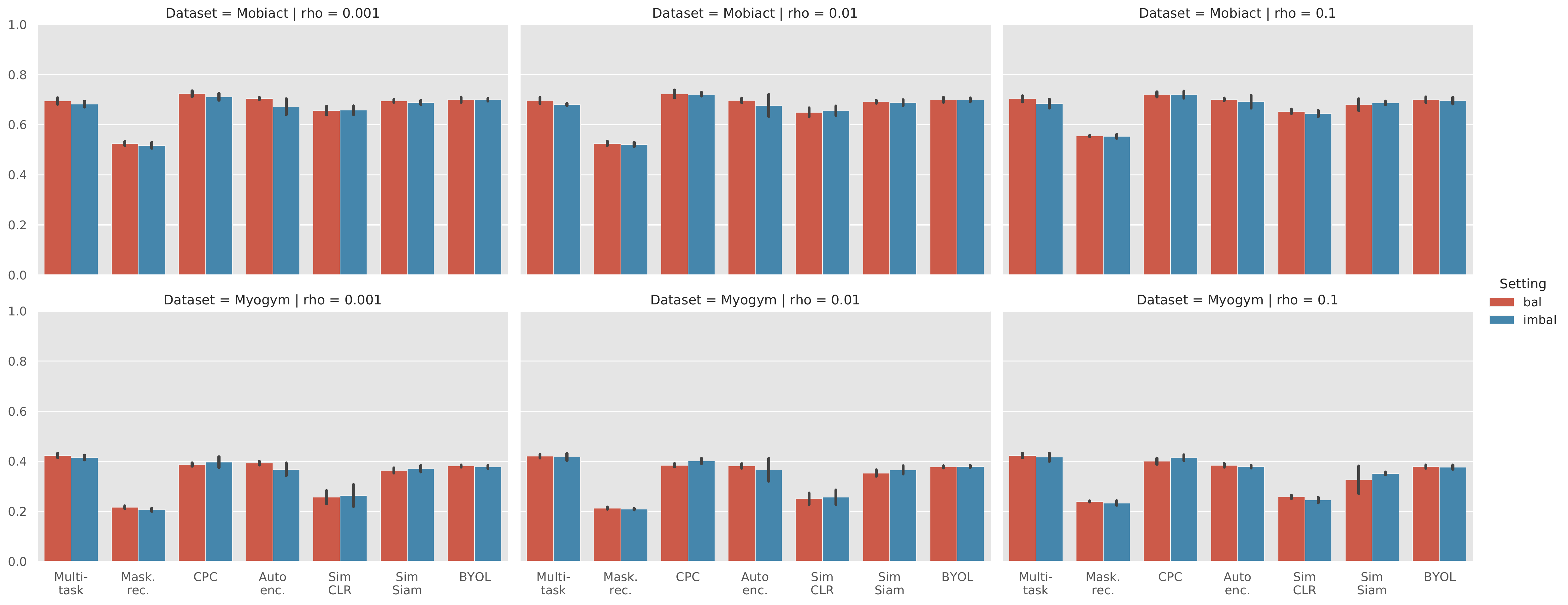}
	\caption{
		Studying the effect of imbalance in the source dataset: we artificially create subsets of the source dataset with varying amounts of imbalance, and utilize them for pre-training.
		The activity recognition is performed per usual with five fold validation on the target datasets, and five randomized runs for both the pre-training and classification.
		We only show the visualization on two target datasets for clarity. 
		The remaining results are a part of the appendix (Fig. \ref{fig:first_source_imbalance} and \ref{fig:second_source_imbalance}).
		We note that all self-supervised methods are robust to class imbalances, with only minor differences in performance between balanced and imbalance source subsets.
	}
	\label{fig:source_class_imb}
	\vspace*{-1em}
\end{figure}

The performance of the self-supervised methods under varying levels of dataset imbalance $\rho$ is shown for two randomly chosen datasets in Fig. \ref{fig:source_class_imb} (to study the performance on all target datasets, refer to Fig. \ref{fig:first_source_imbalance} and \ref{fig:second_source_imbalance} in the Appendix).
For Multi-task self-supervision, Masked reconstruction, CPC, \edit{SimCLR, SimSiam, and BYOL} the performance on both imbalanced and balanced data is nearly identical. 

Autoencoder on the other hand shows \edit{worsening} performance as the source subsets get more imbalanced.
For Mobiact, the difference between the means of the balanced and imbalanced runs is $\sim$0.9\% when $\rho$=0.1 (i.e., the rarest class is 10 times smaller than the most frequenct class). 
This increases to $\sim$2\% and $\sim$3.3\% when $\rho$ is 0.01 and 0.001 respectively. 
Myogym follows a similar trend, with reduction $\in \sim\{0.5\%, 1.6\%, 2.4\%\}$ when $\rho \in \{0.1, 0.01, 0.001\}$.
Thus, the performance gets worse as the pre-training is exposed to increasingly imbalanced subsets.
Further, we note that the standard deviation for the Autoencoder also increases considerably with increasing source imbalance.
This can be explained by considering the `pretext' task for autoencoders -- which involves the reconstruction of input windows. 
The pre-training is dependent on reconstructing the source data, and limited exposure to windows of the rarest class renders the model unable to reconstruct them accurately, thereby resulting in less effective representation learning.
As the rarest class reduces in size with decreasing $\rho$, the downstream activity recognition performance also drops slightly (i.e., $\sim$3.3\% at worst for Mobiact). 
Fig.\ \ref{fig:source_class_imb}, \ref{fig:first_source_imbalance} and \ref{fig:second_source_imbalance} reveal the robustness of the self-supervised approaches to class imbalance in the source dataset.
This is a strong point in favor of utilizing self-supervision, as class imbalance in the source dataset does not strongly affect the downstream activity recognition performance. 
As such, this means that the source dataset need not be artificially balanced for effective performance; rather, the unavailability of even coarse labels (for balancing) is not a crutch for learning useful representations.

\subsubsection{Performance under Availability of Very Limited Annotated Data}
In this criterion, we consider a scenario of vital importance to activity recognition, which is when very limited labeled data are available for classification.
This situation can arise when both collecting and annotating data maybe difficult due to cost and privacy concerns, such as in medical applications.
Alternatively, real-world deployment may allow for acquiring a small labeled dataset from users without significant interruptions (or during device setup).
For such scenarios, it is advantageous to identify methods that can pre-train on a large-scale background dataset, and demonstrate improvements for fine-tuning with limited labels.

\begin{figure}[t]
	\centering
	\includegraphics[width=0.7\textwidth]{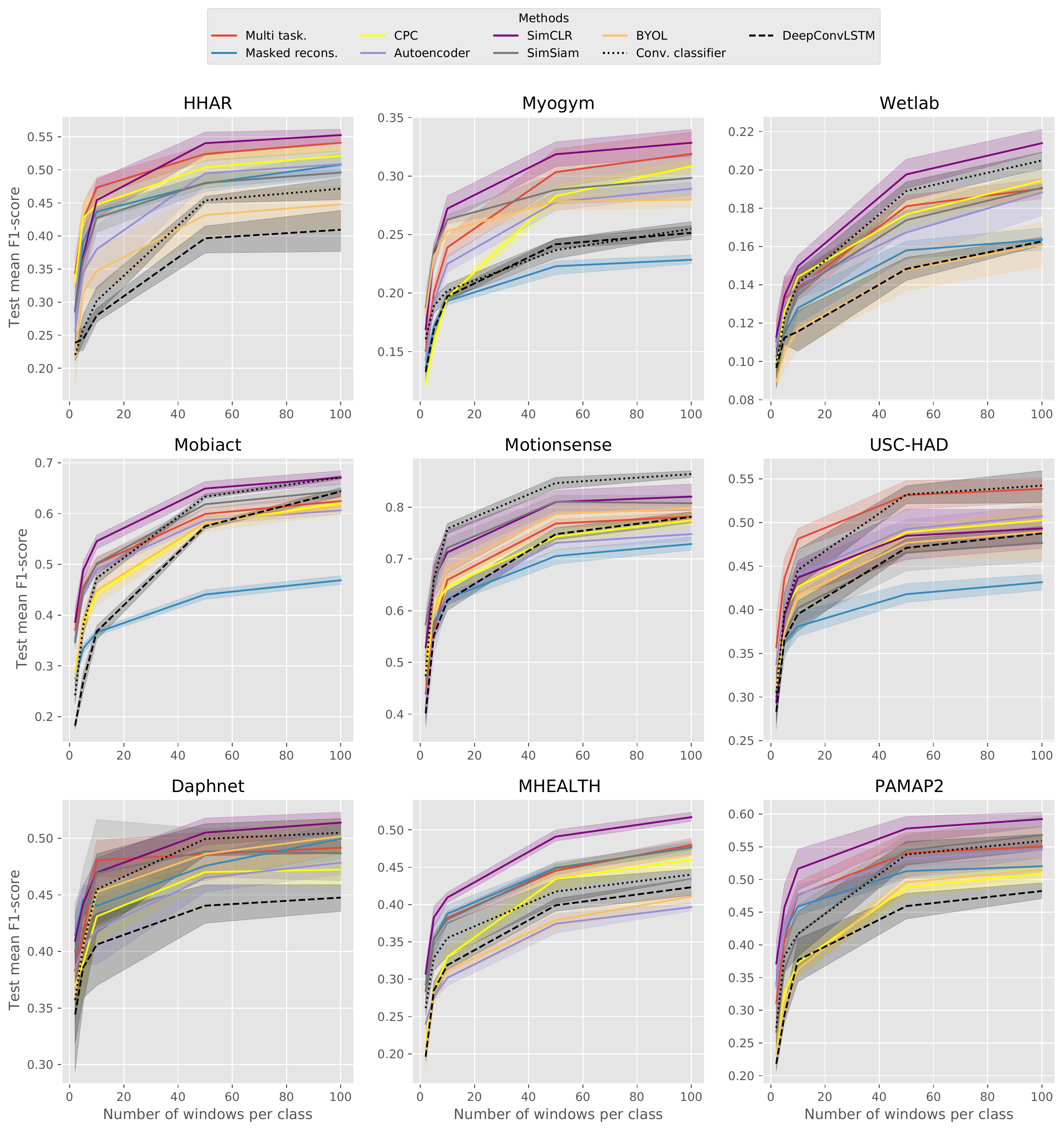}
	\caption{
		Studying the performance of the self-supervised methods when there are very limited data available for fine-tuning: we pre-train the self-supervised approaches on Capture-24 and fine-tune them with limited labeled data, where $\{2, 5, 10, 50, 100\}$ labeled windows per class are available for activity recognition. 
		DeepConvLSTM (in black dashes) is trained end-to-end with the available labeled windows.
		For all methods, the performance across five randomized runs is plotted above.
		We observe significant improvements over DeepConvLSTM for most target datasets.
	}
	\label{fig:limited_labels}
	\vspace*{-1em}
\end{figure}

We compare the performance of the self-supervised approaches against \edit{both} DeepConvLSTM \edit{and Conv. classifier}, which \edit{are} our supervised baselines. 
\edit{
	For clarity, we only show the performance of two best performing end-to-end training approaches.
}
As done previously, we first pre-train with the large-scale Capture-24 dataset and freeze the learned encoder weights. 
Only the classifier layers are optimized using the available labels. 
In contrast, the entire DeepConvLSTM network is trained end-to-end with the annotations.
We utilize the best performing models from Tab. \ref{tab:diff_locs} and \ref{tab:diff_activities} for each self-supervised method and DeepConvLSTM.
For each target dataset fold, we randomly select $\{2, 5, 10, 50, 100\}$ labeled windows per class for training whereas the validation and test splits are untouched.
We compute the mean of the F1-scores obtained across the five folds, and visualize the performance across five randomized runs in Fig. \ref{fig:limited_labels}.

First, we observe that the self-supervised approaches generally outperform DeepConvLSTM (shown in black dashes in Fig. \ref{fig:limited_labels}) by a significant margin. 
For datasets such as Myogym, Wetlab, Daphnet FoG etc., which contain activities that are at least not partially covered in the Capture-24 dataset (see Tab.\ \ref{tab:diff_activities} for reference), we see significant improvements in the test set performance.
This is encouraging as the learned representations can generalize to unseen activities and improve performance even when sufficient labeled data is not available. 
\edit{
	We also note that the Conv. classifier typically performs better than the DeepConvLSTM when training on very limited labels likely due to its relatively small size and simple architecture, leading to less overfitting.
	On target scenarios such as recognizing fine-grained gym activities (Myogym), and for exercise movements (MHEALTH), the self-supervised methods show a substantial improvement over the best supervised technique, i.e., the Conv. classifier. 
	For other datasets, the best performing self-supervised methods still typically outperform the classifier, although the improvements maybe more modest (e.g., in the case of Daphnet FoG dataset and PAMAP2).
}

Analyzing across sensor locations, we notice an interesting trend. 
\edit{
	For HHAR and Myogym, which are wrist-based, as well as MHEALTH, which is leg-based, the self-supervised methods significantly outperform the supervised baseline.
	However, in the case of Motionsense (which is waist-based), the Conv. classifier performs comparably if not better to SimCLR (purple in Fig. \ref{fig:limited_labels}).
	A similar tendency is observed for the other waist datasets such as Mobiact and USC-HAD where the self-supervised methods have a positive impact until 50 labeled windows per class are available.
}
Therefore, self-supervision is more effective when transferring to the wrist-based datasets (which match the source dataset location), or the leg. 
As discussed in Sec. \ref{sec:sensor_positions}, this is likely because many activities covered in the leg-based datasets, such as running, walking, going up and down the stairs have synchronized motion between the wrist and legs. 

\edit{
	In general, SimCLR is the most effective self-supervised method for activity recognition using wearables, across both sensor positions and target activities, whereas Multi-task self-supervision also shows high performance.
	The Conv. classifier is lightweight and thereby more effective for this scenario, oftentimes performing competitively to first pre-training with Capture-24.	
}
The strong performance of the self-supervised methods, even when such limited annotations are available demonstrates the practical value of the pretrain-then-finetune paradigm, for application in scenarios where only small amount of data maybe collected and annotated.

\subsection{Feature Space Characteristics}
In this dimension, we perform an exploration of the feature space itself, in order to understand its properties.
This exploration is done across three criteria: first, we compare the similarity of the learned representations to supervised learning, thereby understanding whether the self-supervised approaches capture similar components of the data as in supervised learning, even though they do not have access to target annotations.
Subsequently, we gauge the linear separability of the representations, assessing whether the learned features enable all downstream tasks or only specific ones.
Finally, we compute the implicit dimensionality of the learned representations for all methods, with the aim of understanding whether the methods fully utilize the representation space or not. 

\subsubsection{Similarity to Supervised Learning}

\begin{figure*}[t]
	\centering
	\includegraphics[width=1\textwidth]{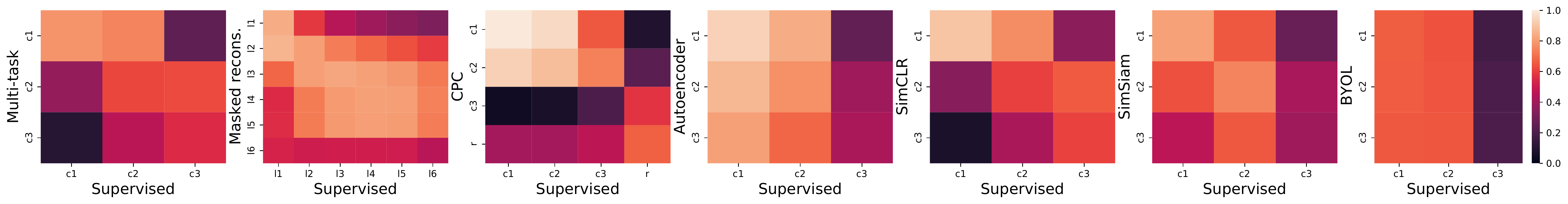}
	\caption{
		Studying the representation similarity to supervised learning: we observe higher similarity between early layers trained with self-supervision and supervised learning. 
	}
	\label{fig:mobi_similarity}
\end{figure*}

The gold-standard for activity recognition generally lies in supervised learning, wherein end-to-end training is performed with the available annotations, resulting in the best possible performance.
However, the introduction of self-supervised approaches has provided an alternative methodology for learning representations, sometimes even outperforming  the supervised counterparts.
As such, an interesting investigation is towards understanding if the supervised and self-supervised methods capture similar components of the data, and if so, which layers are most similar.
We utilize Linear Centered Kernel Alignment (CKA) \cite{kornblith2019similarity}, which returns a value between 0 and 1 (1: identical), in order to obtain a measure of representation similarity.

For this analysis, we take the first fold of target datasets and perform supervised training with a network identical to the self-supervised architecture. 
Subsequently, we perform a forward pass on $1,000$ randomly chosen windows in order to extract the representations at each convolutional layer (i.e., after the layer itself rather than after dropout and ReLU present in the block) of the networks.
Correspondingly, we also extract self-supervised features from previously trained models. 
We perform a pairwise comparison between representations of self-supervised and supervised methods across five random runs, and visualize the mean for Mobiact in Fig. \ref{fig:mobi_similarity}.
For clarity, the visualization for all target datasets has been added to the Appendix, in Fig.\  \ref{fig:all_similarity}. 
A similar exploration was also performed in Multi-task self-supervision \cite{saeed2019multi}, where Singular Vector Canonical Correlation Analysis (SVCCA) \cite{raghu2017svcca} was utilized.
CKA is related to SVCCA, yet it is advantageous as it is able to determine correspondence between layers trained with different random initializations, and more importantly, different feature sizes. 

From Fig. \ref{fig:mobi_similarity}, we observe that the first encoder layer outputs representations that are most similar to supervised learning. 
As in the original paper for CKA \cite{kornblith2019similarity}, the similarity generally reduces as the depth of the encoder layer increases (see the diagonal of the similarity matrices for reference).
This trend is clearer for Masked reconstruction, which contains six Transformer encoder layers.
The diagonal elements until layer $5$ are very similar, especially in the neighborhood around the diagonal.
This indicates that immediately preceding and succeeding layers are more similar than farther layers.
For example, layer $1$ from the pre-trained Masked reconstruction is more similar to the Supervised layer $2$ than layer $6$. 

In the case of CPC, we observe a sharp drop in the similarity after the second convolutional layer, with the features from third layer being considerably dissimilar to preceding layers.
The fourth layer of CPC is the Gated Recurrent Unit \cite{chung2014empirical}, which also learns representations similar to supervised learning.
\edit{
	This is true for BYOL as well, with the final convolutional layer having much lower similarity than the earlier layers.
}
the Autoencoder also follows the general trend of higher similarity in the earlier layers with the first two convolutional layers being very similar.
Interestingly however, the self-supervised representation from the third layer also shows high similarity to the features from the first supervised layer. 

The similarity matrices from Fig.\ \ref{fig:mobi_similarity} and \ref{fig:all_similarity} demonstrate that self-supervised approaches have the capability to learn representations similar to end-to-end training.
More specifically, the earlier layers of the self-supervised network produce very similar representations to supervised learning (a trend also observed in \cite{kornblith2019similarity}).
The similarity of the later layers diverges, likely because the supervised representations capture the more activity-specific characterisics of the data, whereas the self-supervised methods optimize different objectives entirely.

\subsubsection{Linear Separability of the Representations}
\label{sec:linear_sep}
\begin{figure*}
	\centering
	\vspace*{-1em}
	\includegraphics[width=\textwidth]{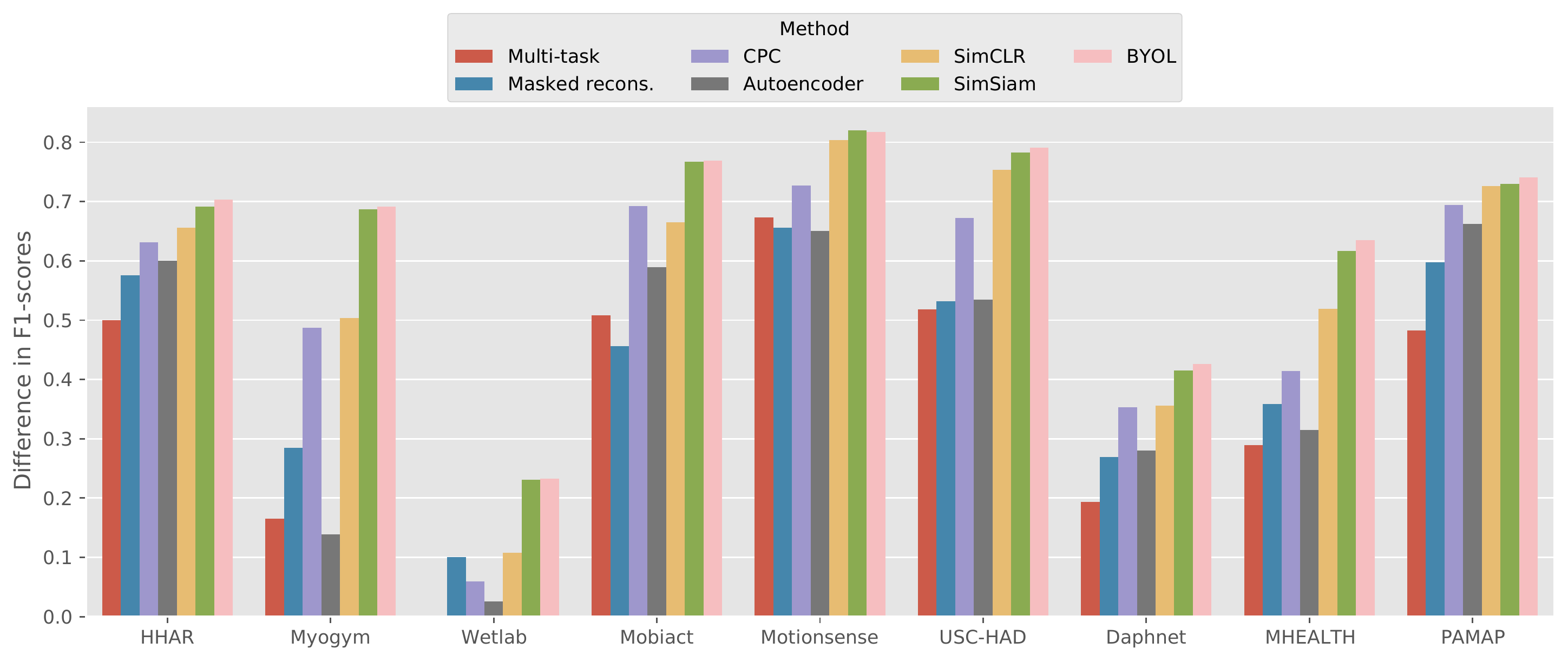}
	\caption{
		Studying the linear separability of the learned representations: we compute the difference in the train set F1-score between utilizing ground truth and random labels using a \textbf{linear} classifier. 
		We observe that Multi-task self-supervision succeeds more in learning generic representations rather than ones specific to downstream tasks. 
	}
	\label{fig:linear_sep}
\end{figure*}

One of the aims of the pretrain-then-finetune paradigm is to learn generic representations from a large body of unlabeled movement data, such that \emph{any} downstream task/scenario can leverage the pre-trained weights.
This criterion studies whether the self-supervised methods learn representations that are only useful for certain downstream tasks or scenarios, or whether they enable all tasks by creating a representation space wherein many labelings of the data are expressible as linear classifiers~\cite{wallace2020extending}.
In order to quantify such separability, we utilize the protocol detailed by Wallace \etal \cite{wallace2020extending} where linear classifiers are trained on the learned encoder weights, but with random labels.
The performance obtained by using the random annotations is contrasted against utilizing the ground truth, and their difference is plotted in Fig.\ \ref{fig:linear_sep}.
Methods that have a smaller difference in the linear classification performance between using ground truth and random labels learn generic feature descriptors rather than capturing representations that are only relevant to some downstream scenarios \cite{wallace2020extending}.
In Fig.\ \ref{fig:linear_sep}, we perform five randomized runs for both the pre-training and the linear classification, and plot the difference in the mean of the train set F1-scores.

We observe that Multi-task self-supervision generally has a lower difference in the performance between ground truth and randomized labels, indicating that it learns more general representations that enable various downstream scenarios.
\edit{
	The contrastive learning methods show the highest difference in the train F1-score thereby learning the least generic features, especially SimCLR, SimSiam, and BYOL.
}


\subsubsection{Implicit Dimensionality of the Representations}
\begin{figure*}
	\centering
	\includegraphics[width=0.4\textwidth]{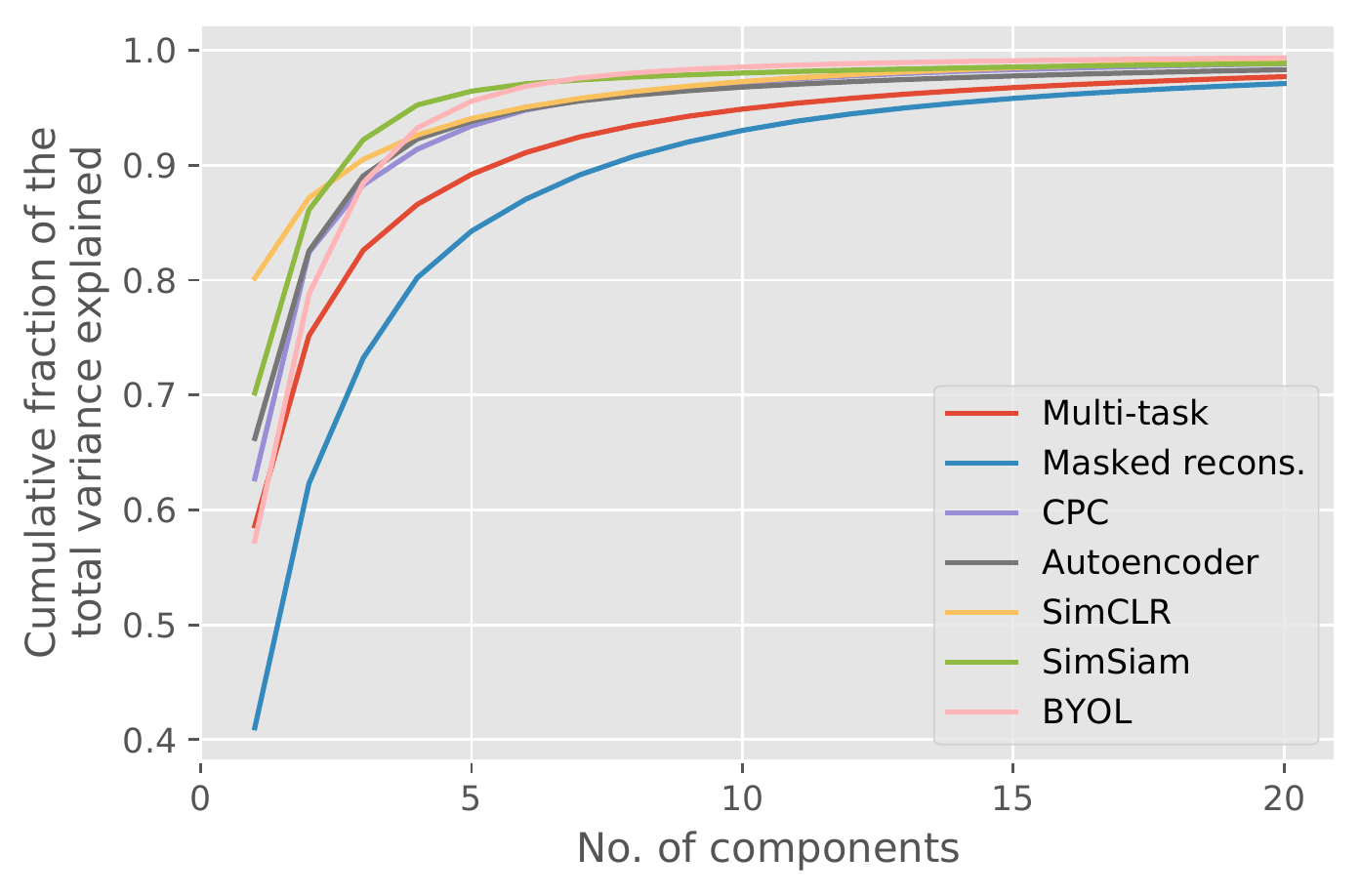}
	\caption{
		Studying the implicit dimensionality of the learned representations: the cumulative fraction of the total variance across the first twenty principal components is averaged across all datasets and visualized.
		We observe that Masked reconstruction has the lowest explained variance thereby indicating \emph{higher} implicit dimensionality.
	}
	\label{fig:imp_dim}
	\vspace*{-1em}
\end{figure*}

The dimensionality of the features used for activity recognition is of importance from a wearables standpoint, as lower feature dimensions result in reduced computational costs and computational time during classification~\cite{haresamudram2019role}.
Approaches that can perform effectively while also resulting in smaller features are more desirable due to the computational savings their offer. 
At the same time, larger representation sizes can benefit from the availability of more training data \cite{kolesnikov2019revisiting}. 
As detailed in \cite{wallace2020extending}, we study the implicit dimensionality of the learned representations through Principal Component Analysis.
The goal of this analysis is to quantify the percentage of the total variance explained by the first $n=20$ principal components.
Higher percentage of the total variance being explanied by the first 20 components (or fewer) indicates relatively \emph{lower} implicit dimensionality.

As in Sec. \ref{sec:linear_sep}, we perform inference using the best performing pre-trained models from Tab.\ \ref{tab:diff_locs} and \ref{tab:diff_activities} on the first fold of the target datasets.
We randomly sample $1,000$ training windows in five randomized runs and extract the representations from the last encoder layer.
In order to make the feature sizes comparable, we utilize the last convolutional layer from CPC for extraction rather than the GRU.
Principal Component Analysis (PCA) is performed on the extracted features for each target dataset, and the cumulative fraction of the explained variance is averaged across the five randomized runs. 
Following \cite{wallace2020extending}, the cumulative fraction of total variance explained in Fig. \ref{fig:imp_dim} is obtained by once again computing the mean across all target datasets. 
For reference, the implicit dimensionality of the learned representations has been visualized individually for each target dataset in Fig.\ \ref{fig:all_imp_dim} in the Appendix.

\edit{
	In Fig.\ \ref{fig:imp_dim}, we see that for the contrastive methods such as CPC, SimCLR, and BYOL, as well as the Autoencoder, around 95\% of the variance can be explained from the first five components itself, which increases to over 98\% when considering 20 components.
	Over 97\% of the variance can be explaind for SimSiam from the first five components, plateauing thereafter.
}

In contrast, for Multi-task self-supervision, less than 90\% of the variance can be explained using five components. 
It is lower still for Masked reconstruction, at around 85\%.
As most of the variance can be explained for CPC\edit{, SimCLR, SimSiam, BYOL,} and Autoencoder using just five components, they have relatively lower implicit dimension. 
On the other hand, Multi-task self-supervision and Masked reconstruction utilize the latent space more effectively and have higher implicit dimensionality.

\section{Discussion}
\label{sec:discussion}
The main goal of this work is to deepen our understanding of self-supervised approaches for human activity recognition using wearables. 
We accomplish this by conducting a large-scale study utilizing the assessment framework, which comprises of a \emph{collection} of criteria.
The first part of this section investigates two choices in our experimental setup -- first, that it is advantageous to utilize the source dataset normalization statistics for target dataset normalization; and, second, simple encoder architectures may not be sufficient for effective representation learning.
The latter part contains a retrospective of the empirical study, wherein we present the benefits and downsides of each self-supervised technique, and summarize the insights gained.
Lastly, we also discuss desirable characteristics for such assessment frameworks, and chart a research agenda for future versions.

\subsection{Applying Source Dataset Normalization to Target Scenario for Activity Recognition}
For all experiments conducted in our study, the pre-processing for the target datasets included applying the normalization means and variances from the source (Capture-24) training set so as to bring the dataset statistics closer.
This practice of applying normalization using the source dataset means and variances is utilized for transfer learning \cite{he2019bag}.
In this section, we study and quantify whether the source dataset means and variances must be applied at all for wearables-based self-supervised transfer learning.

\begin{figure*}
	\centering
	\vspace*{-1em}
	\includegraphics[width=\textwidth]{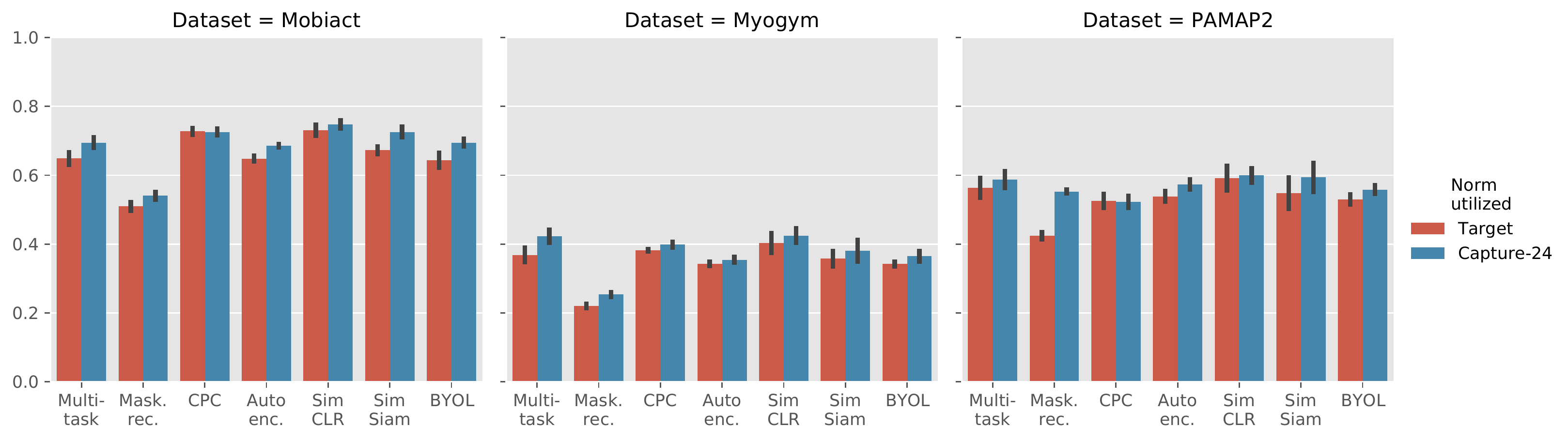}
	\caption{
		Studying whether the source dataset means and variances must be utilized for pre-processing the target datasets during activity recognition: we plot the activity recognition performance when the target datasets are normalized based on the source dataset statistics, or the target data itself.
		There is a significant improvement in performance obtained by utilizing the source dataset (Capture-24) means and variances.
		For clarity, we only show the performance on one dataset per sensor location in this figure. 
		The rest are contained in Fig.\ \ref{fig:all_cap_norm} in the Appendix.
	}
	\label{fig:source_norms}
	\vspace*{-1em}
\end{figure*}

We repeat the setup for activity recognition from Tab.\ \ref{tab:diff_locs} and \ref{tab:diff_activities} while either performing normalization based on the source dataset statistics or based on the target dataset itself.
The test-set F1-score obtained after five-fold validation and five randomized runs is visualized in Fig.\ \ref{fig:source_norms}.
For clarity, we choose one dataset from each sensor location and present the results in the aforementioned figure.

In general, the utilization of the source normalization statistics results in clear improvements over normalizing based on the target data itself.
In the case of Mobiact, Multi-task self-supervision, Masked reconstruction, \edit{Autoencoder, SimSiam, and BYOL} show a considerable increase in performance when the Capture-24 normalization statistics are utilized.
This trend is observed for PAMAP2 as well with Masked reconstruction, \edit{Autoencoder, SimSiam, and BYOL} demonstrating substantial improvements.
For Myogym, Multi-task self-supervision shows the largest difference in performance with the usage of source dataset statistics.
Overall, we note that utilizing the Capture-24 training dataset normalization parameters has the highest impact on Masked Reconstruction and Autoencoder.
This makes sense as the `pretext' for these methods involves reconstructing the input windows (or masked portions of it), and thus applying normalization using the source dataset mean and variances results in similar data statistics. 
Therefore, it is clearly advantageous to utilize the source dataset normalization's means and variances during self-supervised transfer learning. 

\label{sec:source_norms}

\subsection{Is the Performance of Self-Supervision Getting Hampered by (Too) Simple Encoder Architectures?}
\label{sec:bigger_encoder}
\begin{figure*}
	\centering
	\includegraphics[width=0.9\textwidth]{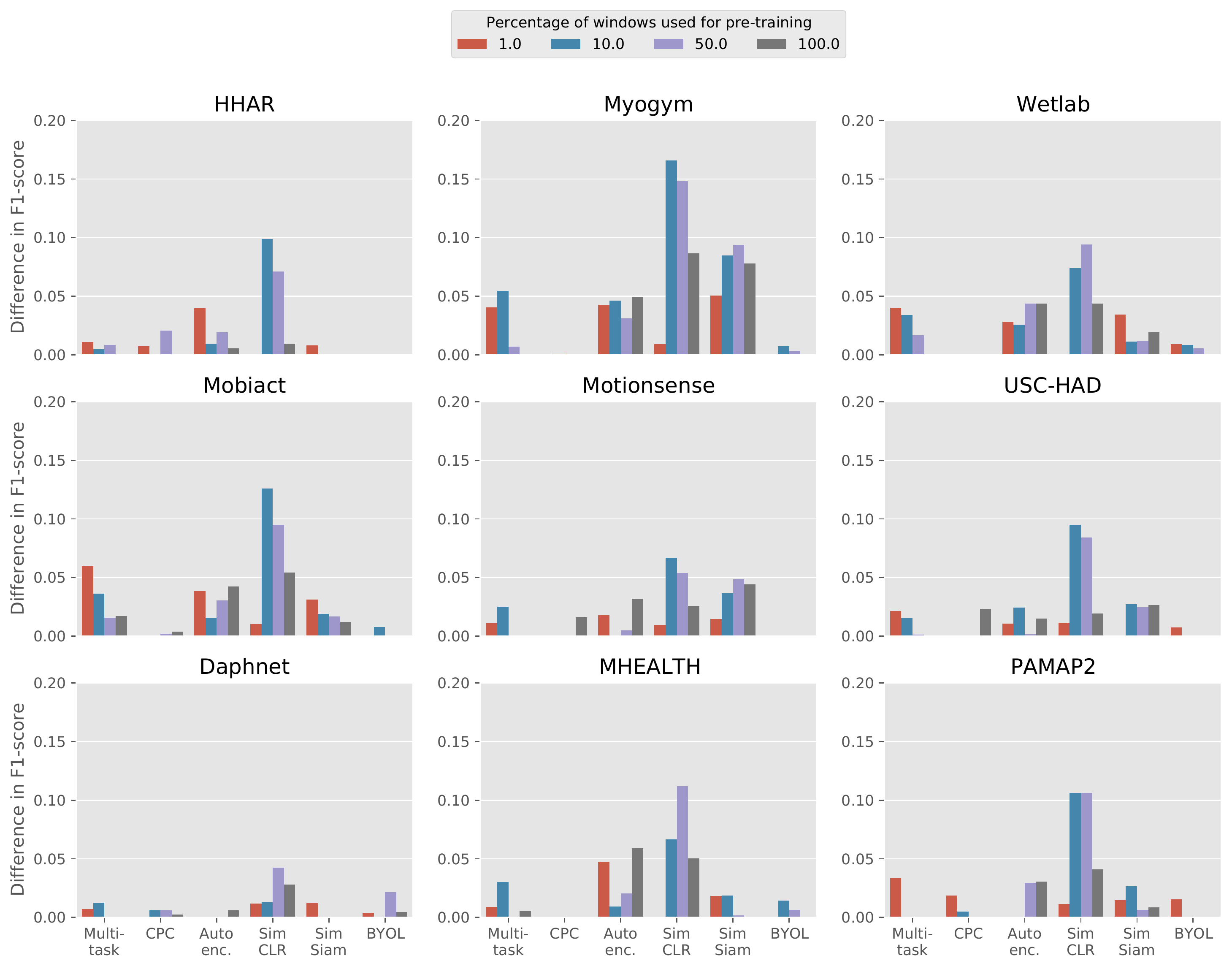}
	\caption{
		Studying the impact of deeper encoder architectures: we increase the depth of the encoder for Multi-task self-supervision, CPC, and Autoencoder by adding an additional convolutional block.
		We plot the difference in the mean of the test set mean F1-scores across five folds of the original and deeper encoder architectures for increasing quantities of pre-training windows.
		Note that a positive difference indicates worsening performance with additional layers. 
		CPC shows no change with the addition of another convolutional block in its encoder. 
	}
	\label{fig:bigger_encoder}
	\vspace*{-1em}
\end{figure*}

In our study of the quantity of the unlabeled source dataset windows required for pre-training (see Fig.\ \ref{fig:perc_windows}), we observed that using only 10\% of the training windows results in comparable if not better downstream activity recognition.
We hypothesized that this was likely due to the simple and shallow encoder architectures employed by these approaches, which comprise generally of three 1D convolutional blocks.
Here, we increase the depth of the encoder by adding a larger convolutional block to the encoders, and study the impact on activity recognition.
For Multi-task self-supervision, we add a fourth convolutional block containing 128 filters and a kernel size of 4, whereas for CPC, the additional convolutional block comprises of 256 filters with the kernel size of 3, matching the previous layers. 
Similarly for the Autoencoder, the new convolutional block contains 256 filters and a kernel size of 9.
We do not perform this analysis for Masked reconstruction as the best overall model during pre-training contains six Transformer \cite{vaswani2017attention} encoder layers and is thus a large encoder.

The pre-training is performed for increasing quantities of randomly sampled pre-training windows, similar to the analysis in Fig.\ \ref{fig:perc_windows}, across \edit{three} randomized runs of sampling the source windows and pre-training, to the five fold validation and activity recognition.
The difference in the mean of the five-fold F1-score for the original encoder architecture and the deeper variant are plotted for all target datasets in Fig. \ref{fig:bigger_encoder}.
In the figure, increasing differences indicate reduced performance resulting from the addition of an extra layer.

We note that there is very limited difference in F1-scores for CPC, for increasing quantities of the source data, indicating that the additional convolutional block did not have any discernable effect on performance.
For both Multi-task self-supervision and Autoencoder, the performance reduces with the addition of another encoder layer, indicating that such naive increase in encoder depth does not show a corresponding improvement in activity recognition, even if more pre-training data are available.
\edit{
	In the case of SimCLR, we see drastic reductions in performance with the large encoder architecture, dropping by over 15\% on Myogym and over 5\% across all target datasets. 
	This trend does not hold for the remaining contrastive methods however, which have a much lower drop in performance.
}
Another likely reason for the reduction in performance is that this experiment utilizes the best overall parameters as determined in Tab.\ \ref{tab:diff_locs} and \ref{tab:diff_activities}.
It is possible that these hyperparameter settings are not ideal for the increased depth of the encoder. 

\subsection{Advantages and Drawbacks of Self-Supervised Approaches}
We have performed a large scale empirical study into assessing the state of the \emph{pretrain-then-finetune} paradigm that is gaining widespread adoption by the wearables community.
In such a paradigm, a large-scale background movement dataset can be utilized to learn effective representations with the goal of enabling a host of diverse downstream scenarios, some of which may have difficulty in collecting data or obtaining annotations. 
Here, we summarize the benefits and drawbacks of each of the evaluated methods from different standpoints as encountered during the course of this study -- such as the training time, computional load, and complexity of the approach, and tabulate them into Tab.\ \ref{tab:summary}. 

\begin{table*}[t]
	\centering
	\vspace*{-1em}
	\caption{
		A brief summary of the advantages and drawbacks of each self-supervised method.
		$\sim$ indicates that the implementation is neither simple nor too complex.
	}
	\vspace*{-1em}
	\begin{tabular}{cP{1.5cm}P{1.5cm}P{2.5cm}P{2.5cm}P{3.3cm}}
		\toprule
		Approach & Simple encoder & Quick training & Low computional load & Simple Implementation & Small hyperparameter space \\ 
		\hline
		Multi-task & $\checkmark$ & $\times$ & $\checkmark$ & $\sim$ & $\checkmark $ \\
		Masked recons. & $\times$ & $\times$ & $\times$ & $\sim$ & $\checkmark $ \\
		CPC & $\checkmark$ & $\times$ & $\times$ & $\times$ & $\times $ \\
		Autoencoder & $\checkmark$ & $\checkmark$ & $\checkmark$ & $\checkmark $ & $\checkmark$ \\
		\edit{SimCLR} & \edit{$\checkmark$} & \edit{$\checkmark$} & \edit{$\checkmark$} & \edit{$\sim$} & \edit{$ \checkmark $} \\
		\edit{SimSiam} & \edit{$\checkmark$} & \edit{$\checkmark$} & \edit{$\checkmark$} & \edit{$\sim$} & \edit{$ \checkmark $} \\
		\edit{BYOL} & \edit{$\checkmark$} & \edit{$\checkmark$} & \edit{$\checkmark$} & \edit{$\sim$} & \edit{$\checkmark $} \\ 
		\bottomrule
	\end{tabular}
	\label{tab:summary}
	\vspace*{-1em}
\end{table*}

Multi-task self-supervision \cite{saeed2019multi} jumpstarted the investigation into the applicability of self-supervision for human activity recognition. 
The main advantage of utilizing this approach lies in its effectiveness in learning representations that are applicable for a variety of downstream scenarios, as evidenced by its overall superior performance in Tab.\ \ref{tab:diff_locs} and \ref{tab:diff_activities}.
The transformation techniques applied in this method are well designed for learning representations for accelerometer data. 
The encoder architecture is simple, resulting in small GPU memory footprint during pre-training as well as fine-tuning.
The hyperparameter space we utilized is also small, consisting primarily of the learning rate and L2 regularization.
However, there is a host of hyperparameters regarding the signal transformations which can be tuned for improved performance (we do not perform that exploration in this study).
One drawback includes its relatively poor performance on multi-sensor datasets \cite{haresamudram2020masked, haresamudram2021contrastive} as the data transformations are specifically designed for accelerometry and cannot be trivially applied to other sensors, for example, gyroscopes and magnetometers, which are typically part of the IMU setups.

Autoencoders are one of the most well established unsupervised learning approaches. 
The primary benefit of utilizing autoencoders lies in their simplicity, with the encoder and decoder architectures being mirror images of each other, and the objective corresponding to simply reconstructing the original signal. 
Therefore, they are easy to setup and typically one of the baselines against which self-supervised methods are compared. 
Additionally, we observe in our study that Autoencoders often perform comparably to more sophisticated methods such as Multi-task self-supervision or CPC.
Given the simplicity of the architecture and the objective, the training time is also significantly shorter relative to other methods, while also being less stressful from a computational perspective. 
The hyperparameter space is small as well, covering the learning rate, L2 regularization and the convolutional kernel size.
The disadvantage of utilizing autoencoders is that it may not result the best possible unsupervised performance. 
Additionally, care must be taken to not learn an identity function, thereby resulting in less useful representations.

Contrastive Predictive Coding (CPC) \cite{haresamudram2021contrastive} learns representations by predicting multiple future timesteps of sensory data and optimizing via the InfoNCE \cite{van2018representation} loss. 
The encoder architecture is simple, and the approach generally performs well across target scenarios. 
A strength of this method lies in its applicability and superior performance on multi-sensor data~\cite{haresamudram2021contrastive}, and in scenarios where transition-style activities are present (for example, getting in and out of a car, as in Mobiact). 
A downside to this approach is that it requires considerable parameter tuning, including learning rate and weight decay,  future prediction horizon, and  batch size.
Additionally, care must be taken for encoder design and sampling negatives to not produce trivial pre-training, resulting in useless representations. 
The training time is considerable, and extends particularly with increasing prediction horizons.

\edit{
	SimCLR \cite{chen2020simple, tang2020exploring} learns representations by maximizing the agreement different augmented \emph{views} of the same input window via the NT-Xent (normalized temperature scaled cross entropy)  loss \cite{chen2020simple}. 
	The primary advantages of this method include the simplicity of its design, which does not involve specialized architectures or a memory bank.
	In our case as well, we utilized a simple 1D convolutional encoder identical to \cite{saeed2019multi}, as detailed in \cite{tang2020exploring} which was an early adaptation of the technique to sensor data (Motionsense).
	The training itself is relatively fast, while the hyper-parameter space is not very large, comprising of learning rate, weight decay, and the batch size.
	While it is computationally light as well, the setup necessitates large batch sizes, as the negative pairs for contrastive learning consist of all other augmented windows in the batch.
	As mentioned in the original paper \cite{chen2020simple}, the choice of augmentations has a large impact on the pre-training quality.
	In this paper, we utilize the augmentations detailed in \cite{tang2020exploring} although we do not optimize on how to chain multiple transformations as it would significantly increase the pre-training search space.
	One potential downside could be its effectiveness on multi-sensory datasets due to its heavy reliance on augmentations, as the established sensor augmentations were designed specifically for accelerometry (only). 
	Overall, it is the most effective representation learning method evaluated in this study while also being methodologically simple and easy to understand.
}

\edit{
	BYOL \cite{grill2020bootstrap} presents a new approach towards self-supervised learning by rejecting the necessity of having negative samples by instead utilizing two separate networks that interact and learn from one another.
	As with SimCLR and SimSiam, the encoder network is based on simple 1D convolutions and is identical to \cite{saeed2019multi}.
	The positives for this technique include its robustness to the choice of augmentations (in direct contrast to SimCLR), simple setup and architecture, and its general effectiveness across a variety of downstream scenarios. 
	This method also trains quickly and is not computationally heavy, while also keeping the pre-training hyper-parameter space relatively small. 
	Altogether, this approach is particularly unique and interesting in its formulation as it features two networks that interact with each other. 
}

\edit{
	SimSiam \cite{chen2021exploring} produces surprisingly effective representations with a simple siamese setup. 
	The main advantages of this method include its simplicity of design, where neither negative pairs nor a momentum encoder or large batch sizes are required for effective learning. 
	As per the original paper \cite{chen2021exploring}, it can be thought of as `BYOL without the momentum encoder' and `SimCLR without the negative pairs', thereby acting as the common connection point between several existing methods. 
	The model architecture is simple with low training times and a small hyperparameter space. 
	The downside to this approach is that it can collapse to trivial solutions, but the stop gradient operation can generally counteract such scenarios.
	As in the case of SimCLR, the augmentation transformations can further be studied for improved performance.
}

Lastly, Masked reconstruction \cite{haresamudram2020masked} is trained by reconstructing only masked out timesteps of the window.
Therefore, a strong point in favor of this approach is that the architecture setup and the training objective are straightforward.
However, as seen in this study, the approach does not generally perform comparably to other self-supervised approaches for learning representations on a single accelerometer. 
Additionally, the Transformer \cite{vaswani2017attention} encoder architecture is large and computationally intensive, and has the longest training time by far in this study. 
The hyperparameter space contains a few parameters, including the number of warmup steps in the learning schedule, the number of Transformer encoder layers and heads, and the percentage of time-steps to be masked. 
In the original paper \cite{haresamudram2020masked} however, this method outperforms supervised and unsupervised baselines on two out of four benchmark datasets which contain data from both an accelerometer and a gyroscope.

Overall, the aim of discussing the pros and cons of each approach is to bring to light other viewpoints regarding these methods that may not immediately be obvious by only considering the resulting activity recognition performance.
Doing so empowers practitioners in the community in making more informed judgement for applying these methods to diverse target scenarios. 

\vspace*{-.5em}
\subsection{The Necessity for an Evolving Benchmark and a Call to Action}
Over the last few years, the human activity recognition community has seen a growing interest into studying and developing self-supervised representation learning methods.
These approaches strike at the heart of the small labeled dataset problem prevalent in sensor datasets, and therefore present the `pretrain-then-finetune' paradigm shift.
In moving away from designing end-to-end architectures that rely solely on the availability of annotated datasets, these methods exploit potentially large-scale unlabeled data to pre-train generic representations that can be subsequently fine-tuned to the specific activities of interest, positively impacting scenarios where data collection and annotation are beset with cost and privacy challenges.
Techniques towards understanding their performance has so far been limited to one view -- wherein the source and target conditions such as sensor locations and activities are similar. 

In this work, we follow the philosophy espoused by natural language processing benchmarks such as GLUE~\cite{wang2018glue} and GEM~\cite{gehrmann2021gem}, which argues that evaluation on a \emph{suite} of tasks is necessary for a well-rounded understanding of model performance.
More specifically, our assessment is similar to GEM, wherein the goal is to foster a nuanced analysis and discussion of performance by the community, unencumbered by relentless pursuit of hill climbing and leaderboards.
As such, this work presents the $v1.0$ into evaluating the current state-of-the-field. 

With the evolution of the field and introduction of increasingly effective and complex techniques, the evaluation and benchmarks need to be updated in tandem with contributions from the community. 
The solved tasks need to be replaced with more challenging and relevant ones, addressing discovered flaws in the established experimental setups~\cite{gehrmann2021gem}. 
Therefore, this work presents a call-to-action for the wearables community as a whole so that we can collectively update and evolve the assessment framework alongside the progress of the field, so that future versions of the framework are large-scale, decentralized community efforts.

\subsection{Lessons Learned and Insights Gained}
\label{sec:lessons_learned}
Through the course of conducting this large-scale empirical study, we have discovered several insights and learned lessons into the behavior of self-supervised methods under diverse conditions. 
Here, we summarize them with brevity for easy reference: 
\begin{enumerate}
	\item \edit{With the aid of more powerful classifiers (such as the MLP-classifier utilized in our study), self-supervised methods are comparable if not better in scenarios wherein the target sensor location does not match the source location, with transfer from the wrist$\rightarrow$leg being more successful than wrist$\rightarrow$waist.} 
	
	\item The approaches have the capability to generalize well to potentially unseen activity classes during downstream recognition, including fine-grained gym exercises such as curls and rows, and medical applications involving recognizing the freeze of gait symptoms in Parkinson`s disease.
	
	\item The target dataset sampling rate must match the source conditions for optimal performance.
	
	\item The techniques are data efficient and perform comparably while only utilizing a small fraction of the available data -- both from a participant as well as quantity of data standpoints -- thereby reducing the computational burden. 
	By and large, a small yet diverse dataset (in terms of participants) can result in better recognition than collecting lots of data from a few participants.
	
	\item Self-supervised learning is surprisingly robust to even extreme imbalances in the source dataset\edit{, with  minimal reduction in performance relative to balanced data}.
	
	\item These approaches have strong practical value, exhibiting significant improvements in scenarios where labeled data is at a premium. 
	
	\item Early layers of the learned encoders show high similarity to supervised learning, whereas the features diverge deeper into the network. 
	Also, layers closer to each other are more similar than those farther apart.
	
	\item Multi-task self-supervision succeeds in learning the most generic features of wearable data, and thereby enables wide variety of downstream tasks.
	
	\item Masked reconstruction has the highest implicit dimensionality and thus more efficiently makes use of the learned representation space.
	
	\item Utilizing the means and variances of the source dataset normalization on the target dataset results in considerable performance gains.
\end{enumerate}

\section{Summary and Conclusion}
The field of human activity recognition using wearables (HAR) is currently undergoing a transition regarding the de-facto standard for how to model and recognize activities.
The community has begun to move away from designing complex end-to-end architectures that require large quantities of annotated data, and rather focuses on leveraging--easier to collect--unlabeled sensor data to derive generic data representations that are then subsequently fine-tuned to the specific downstream scenarios.
Recently, especially self-supervised approaches have demonstrated great promise towards more robust modeling and thus improved activity recognition performance but also opened up new HAR application domains.

In this work, we aimed at assessing the state-of-the-field of self-supervision in HAR research.
We conducted a systematic, large-scale "stress test" of these contemporary methods under a variety of conditions.
We formulated and introduced an assessment framework that comprises of three dimensions, each containing three criteria, which collectively shed light on different aspects of the models.
The first dimension studies the robustness of these approaches when the source and target conditions are not similar, whereas the impact of the dataset characteristics such as class imbalance and quantity of data available is investigated in the second dimension.
The third dimension explores the learned representation space, probing the linear separability, implicit dimensionality, and similarity to supervised learning. 
Put together, these dimensions allow for a multi-faceted evaluation of the model performance, and an inventory of the state of self-supervised human activity recognition research.

Using this assessment framework, we performed our evaluation study on \edit{single accelerometer data from} a curated collection of nine benchmark datasets, which are diverse in terms of the sensor locations, activities under study, data size, and the number of participants.
The study revealed a collection of insights regarding model behavior, for example, the robustness of self-supervised methods towards class imbalances, and the ability to produce significant improvements over supervised learning when there is a scarcity of annotations. 
These results and findings are encouraging as they indicate the broad applicability of these self-supervised methods in wide-ranging target scenarios, and affirm the capability of these techniques for empowering human activity recognition in applications where data both collection and annotation may be challenging to accomplish.
\edit{
	As we only utilize data from a single accelerometer for the assessment, the current version of the framework does not take into consideration issues typically arise in multi-sensor datasets, such as variation in data distribution across devices, synchronization etc.
}

Our assessment framework should be considered as $v1.0$ that allowed us develop a deeper understanding of self-supervised learning methods in the field of wearables-based human activity recognition (HAR).
We call upon the research community for action to collectively contribute towards evolving this framework and using it as a standard for model development and evaluation in our field.
We aim at releasing the framework in an accessible form that will allow researchers and practitioners to employ it for their model development and deployment activities, which--overall--will allow the community to push the state-of-the-art in HAR.

\section*{Acknowledgments}
This work was supported by NSF IIS-2112633. 
Fig \ref{fig:self_sup} utilized icons created by Anuar Zhumaev and Teewara soontom from the Noun Project.



\bibliographystyle{ACM-Reference-Format}
\bibliography{refs}

\appendix
\section{Appendix}

\subsection{Architectures, hyperparameters, and implementation details}
\label{sec:imp_details}
\subsubsection{Multi-task Self-supervision}
The encoder architecture is identical to the original paper \cite{saeed2019multi}, which contained three 1D convolutional layers, having 32, 64 and 96 filters, with a kernel size of 24, 18 and 8, respectively.
Each convolutional layer is followed by the ReLU activation function \cite{nair2010rectified} as well as Dropout \cite{srivastava2014dropout} with p=0.1. 
Global max pooling is applied after the last convolutional layer, thereby forming the encoder network. 
During pre-training, we perform a grid search over the learning rates $\in \{1e-4, 3e-4, 5e-4\}$ and L2 reguralization $\in \{1e-4, 3e-4, 5e-4\}$.
For classification, the parameters include learning rates $\in \{1e-4, 3e-4, 5e-4\}$ and L2 reguralization $\in \{1e-4, 3e-4, 5e-4\}$.

\subsubsection{Masked Reconstruction}
Masked reconstruction is trained using a Transformer encoder \cite{vaswani2017attention} as detailed in the original paper \cite{haresamudram2020masked}.
The raw accelerometer data is transformed to 128 dimensional embeddings using a 1D convolutional layer. 
In order to inject a sense of time (or sequence) to the Transformer encoder, fixed sinusoidal embeddings are utilized and added to the embeddings for input to the encoder. 
The pre-training parameter space includes the number the layers $\in \{2, 3, 4, 5, 6\}$, the number of warmup steps in the Noam optimizer schedule $\in \{20k, 40k, 60k, 80k, 100k\}$, percentage of time steps masked $\in \{10, 20, 30, 40, 50, 60, 70\}\%$.
The number of heads is set to 8, and the classifier learning rate is tuned $\in \{1e-4, 5e-4, 1e-5\}$ with the L2 regularization~$\in \{0.0, 1e-4, 1e-5\}$.

\subsubsection{CPC}
We utilize an identical setup to \cite{haresamudram2021contrastive}, which contains a convolutional encoder as well as a Gated Recurrent Unit (GRU) network \cite{chung2014empirical}.
The convolutional encoder comprises of 3 blocks, with each block containing a 1D convolutional network with reflect padding, followed by the ReLU \cite{nair2010rectified} activation function and Dropout \cite{srivastava2014dropout} with p=0.2.
The blocks contains 32, 64, and 128 filters respectively, and the GRU has a size of 256 units, 2 layers and Dropout with p=0.2.
Pre-training is performed over the following params: learning rate  $\in \{1e-3, 5e-4, 1e-4\}$, L2 regularization  $\in \{0.0, 1e-4, 1e-5\}$, kernel size $\in \{3, 5\}$, and batch size $\in \{64, 128, 256\}$.
The classification utilizes learning rates $\in \{1e-4, 5e-4, 1e-5\}$ and L2 regularization $\in \{0.0, 1e-4, 1e-5\}$.

\subsubsection{Autoencoder}
The encoder for this method is identical to the convolutional encoder of CPC \cite{haresamudram2021contrastive} (detailed above), and contains three convolutional blocks.
The decoder is the mirror opposite of the encoder, also consisting of three convolutional blocks but with reducing number of filters, 128, 64, and 32, respectively.
The pre-training parameters include learning rate $\in \{1e-3, 5e-4, 1e-4\}$, weight decay $\in \{0.0, 1e-4, 1e-5\}$, and kernel size $\in \{3, 5, 7, 9, 11\}$.
For classification, the learning rates $\in \{1e-4, 5e-4, 1e-5\}$ and L2 regularization $\in \{0.0, 1e-4, 1e-5\}$.

\subsubsection{SimCLR}
\edit{
	As in \cite{tang2020exploring}, the convolutional encoder for this method is identical to the one used in Multi-task self-supervision \cite{saeed2019multi} and comprises of three 1D convolutional blocks of 23, 64, and 96 filters, respectively. 
	The corresponding kernel sizes are 24, 16, and 8, with the ReLU activation function and dropout in between. 
	For pre-training, the parameters comprise of learning rate $\in \{1e-2, 1e-3, 5e-3, 1e-4\}$, weight decay $\in \{0.0, 1e-4, 1e-5\}$, batch size $\in \{1024, 2048, 4096\}$. 
	Following \cite{tang2020exploring}, we utilize the SGD optimizer instead with a momentum of $0.9$. 
	As in the original paper \cite{chen2020simple}, a cosine learning schedule is employed for pre-training and the parameters are updated with NT-XentLoss.
	The classification is performed using the learning rates $\in \{1e-4, 5e-4, 1e-5\}$ and L2 regularization $\in \{0.0, 1e-4, 1e-5\}$.
	The projection head is a multi-layer perceptron (MLP) containing three linear layers of 256, 128 and 50 units respectively (similar to the setup from \cite{tang2020exploring}) with the ReLU function applied in-between.
	Additionally, the augmentation transformations are also identical to \cite{tang2020exploring}, with one difference: we do not chain multiple transformations and apply them to each batch.
}

\subsubsection{SimSiam}
\edit{
	The backbone is identical to \cite{saeed2019multi} and \cite{tang2020exploring} and the pre-training is performed over learning rate $\in \{1e-2, 5e-2, 1e-3, 5e-3, 1e-4, 5e-4, 1e-5\}$, weight decay $\in \{0.0, 1e-4, 1e-5\}$, batch size $\in \{128, 256, 512\}$. 
	Based on \cite{chen2021exploring}, we use the SGD optimizer with a momentum of $0.9$, with a cosine learning schedule . 
	As in the original paper \cite{chen2020simple}, a cosine learning schedule is employed for pre-training and the parameters are updated with a symmetric cosine similarity based loss.
	The classification is performed using the learning rates $\in \{1e-4, 5e-4, 1e-5\}$ and L2 regularization $\in \{0.0, 1e-4, 1e-5\}$.
	The projection head comprises of three linear layers of 128, 128, and 96 units, respectively, with batch normalization \cite{ioffe2015batch} and ReLU after each layer.
	For the prediction head, we use a smaller network (as per \cite{chen2021exploring}) containing two linear layers of 64 and 96 units along with batch normalization and ReLU being applied after the first layer.
	As with SimCLR (detailed above), we apply two transformations from the collection defined in \cite{tang2020exploring} for every window in the batch.
}

\subsubsection{BYOL}
\edit{
	As with the two prior approaches, the backbone is a lightweight convolutional encoder as defined in \cite{saeed2019multi}. 
	The optimization is performed with a SGD optimizer with a momentum of 0.9 and a cosine learning rate schedule, with learning rate $\in \{1e-2, 5e-2, 1e-3, 5e-3, 1e-4, 5e-4, 1e-5\}$, weight decay $\in \{0.0, 1e-4, 1e-5\}$, batch size $\in \{512, 1024, 2048, 4096\}$. 
	Again, the classifier parameters include learning rates $\in \{1e-4, 5e-4, 1e-5\}$ and L2 regularization $\in \{0.0, 1e-4, 1e-5\}$.
	The projection and prediction heads both contain two linear layers of $\{256, 64\}$ and $\{128, 64\}$ units respectively with batch normalization and ReLU between successive layers.
}

\subsubsection{Linear Evaluation}
\edit{	
	\label{sec:linear_classifier}
	It comprises of a single fully-connected layer without any activation functions and the number of units depends on the number of classes present in the target dataset.
}

\subsubsection{MLP Classifier}
\label{sec:mlp_classifier}
All self-supervised methods are evaluated on a common backend network, identical to the classifier described in \cite{haresamudram2021contrastive, haresamudram2020masked}.
After the pre-training is complete, the encoder weights are frozen and only the classifier network is updated via the cross entropy loss.
It consists of three linear layers of 256, 128 and $num\_classes$ units respectively. 
Between each layer, batch normalization \cite{ioffe2015batch}, the ReLU activation function, and Dropout with p=0.2 are applied consecutively. 

\subsubsection{DeepConvLSTM}
\label{sec:deepconvlstm}
We implement the DeepConvLSTM architecture detailed in \cite{ordonez2016deep}, which contains four 2D convolutional layers containing 64 filters and kernel size of $5\times1$. 
This is followed by a a LSTM network of 2 layers and 128 units, followed by a linear classifier layer. 
The parameter search is performed over learning rate $\in \{1e-3, 5e-4, 1e-4\}$, and weight decay $\in \{0.0, 1e-4, 1e-5\}$.

\subsubsection{LSTM and GRU Classifiers}
\label{sec:lstm_classifier}
\edit{
	We  use a single unidirectional LSTM or GRU layer of $128$ units, followed by a dropout of $p=0.2$ and a fully connected softmax layer (the number of units are dependent on the number of target activities). 
	The training utilizes learning rate $\in \{1e-3, 5e-4, 1e-4\}$, and L2 regularization $\in \{0.0, 1e-4, 1e-5\}$. 
}

\subsubsection{Convolutional Classifier}
\label{sec:conv_classifier}
\edit{
	As many of the unsupervised baselines consist of a convolutional encoder identical to architecture detailed in \cite{saeed2019multi}, we study its performance for end-to-end training.
	We pass the output from the convolutional encoder through a fully connected layer with the number of units equalling the number of target activities.
	The training is performed with learning rate $\in \{1e-3, 5e-4, 1e-4\}$, and L2 regularization $\in \{0.0, 1e-4, 1e-5\}$. 
	As with DeepConvLSTM, the learning rate is reduced by a factor of 0.8 every 10 epochs.  
}

\begin{table}[h]
	\centering
	\caption{Best performing hyperparameters for Multi-task self-supervision for each target dataset. }
	\begin{tabular}{|c|c|c|c|c|c|c|}
		\hline
		Dataset & lr & L2 reg. & class. lr & class. L2 reg. & F1-score (mean) & F1-score (std) \\ \hline
		HHAR & 0.0005 & 0.0001 & 0.0001 & 0.00001 & 58.1 & 1.06 \\ \hline
		Myogym & 0.0005 & 0.00001 & 0.0005 & 0.0001 & 39.89 & 0.98 \\ \hline
		Wetlab & 0.001 & 0.0001 & 0.0005 & 0 & 24.16 & 0.48 \\ \hline
		Mobiact & 0.001 & 0.00001 & 0.0005 & 0 & 72.91 & 0.99 \\ \hline
		Motionsense & 0.001 & 0 & 0.0001 & 0.00001 & 84.74 & 1.14 \\ \hline
		USC-HAD & 0.0005 & 0.00001 & 0.0005 & 0.0001 & 51.37 & 2.43 \\ \hline
		Daphnet & 0.001 & 0.0001 & 0.0005 & 0 & 51.16 & 1 \\ \hline
		MHEALTH & 0.0005 & 0.00001 & 0.0005 & 0.0001 & 45.49 & 1.27 \\ \hline
		PAMAP2 & 0.001 & 0 & 0.0001 & 0.00001 & 52.24 & 1.98 \\ \hline
	\end{tabular}
\label{tab:multi_params}
\end{table}

\begin{table}[h]
	\centering
	\caption{Best performing hyperparameters for Masked Reconstruction for each target dataset. }
	\begin{tabular}{|c|c|c|c|c|c|c|c|c|}
		\hline
		Dataset & head & layers & mask\% & warmup & class. lr & class. L2 reg. & F1-score (mean) & F1-score (std) \\ \hline
		HHAR & 8 & 6 & 10 & 80000 & 0.0005 & 0.0001 & 55.04 & 2.58 \\ \hline
		Myogym & 8 & 3 & 40 & 80000 & 0.0005 & 0.0001 & 25.29 & 0.68 \\ \hline
		Wetlab & 8 & 6 & 10 & 80000 & 0.0005 & 0.0001 & 21.23 & 0.31 \\ \hline
		Mobiact & 8 & 6 & 70 & 60000 & 0.0005 & 0 & 54.17 & 1.38 \\ \hline
		Motionsense & 8 & 4 & 50 & 60000 & 0.0001 & 0 & 75.72 & 1.88 \\ \hline
		USC-HAD & 8 & 4 & 70 & 80000 & 0.0005 & 0.0001 & 45.09 & 0.92 \\ \hline
		Daphnet & 8 & 4 & 50 & 60000 & 0.0005 & 0.0001 & 52.51 & 1.01 \\ \hline
		MHEALTH & 8 & 6 & 40 & 40000 & 0.0005 & 0.00001 & 47.04 & 0.61 \\ \hline
		PAMAP2 & 8 & 5 & 60 & 60000 & 0.0005 & 0 & 55.12 & 0.96 \\ \hline
	\end{tabular}
\label{tab:masked_params}
\end{table}

\begin{table}[]
	\centering
	\caption{Best performing hyperparameters for CPC for each target dataset. }
	\begin{tabular}{|c|c|c|c|c|c|c|c|}
		\hline
		Dataset & k & lr & L2 reg. & class. lr & class. L2 reg. & F1-score (mean) & F1-score (std) \\ \hline
		HHAR & 48 & 0.0005 & 0.0001 & 0.0001 & 0.00001 & 58.1 & 1.06 \\ \hline
		Myogym & 48 & 0.0005 & 0.00001 & 0.0005 & 0.0001 & 39.89 & 0.98 \\ \hline
		Wetlab & 48 & 0.001 & 0.0001 & 0.0005 & 0 & 24.16 & 0.48 \\ \hline
		Mobiact & 64 & 0.001 & 0.00001 & 0.0005 & 0 & 72.91 & 0.99 \\ \hline
		Motionsense & 32 & 0.001 & 0 & 0.0001 & 0.00001 & 84.74 & 1.14 \\ \hline
		USC-HAD & 48 & 0.0005 & 0.00001 & 0.0005 & 0.0001 & 51.37 & 2.43 \\ \hline
		Daphnet & 48 & 0.001 & 0.0001 & 0.0005 & 0 & 51.16 & 1 \\ \hline
		MHEALTH & 48 & 0.0005 & 0.00001 & 0.0005 & 0.0001 & 45.49 & 1.27 \\ \hline
		PAMAP2 & 32 & 0.001 & 0 & 0.0001 & 0.00001 & 52.24 & 1.98 \\ \hline
	\end{tabular}
\label{tab:cpc_params}
\end{table}

\begin{table}[]
	\centering
	\caption{Best performing hyperparameters for Autoencoder for each target dataset. }
	\begin{tabular}{|c|c|c|c|c|c|c|c|}
		\hline
		Dataset & kernel\_size & lr & wd & class. lr & class. L2 reg. & F1-score (mean) & F1-score (std) \\ \hline
		HHAR & 11 & 0.001 & 0 & 0.0001 & 0.00001 & 54.25 & 2.04 \\ \hline
		Myogym & 9 & 0.001 & 0.00001 & 0.0005 & 0.00001 & 35.45 & 0.49 \\ \hline
		Wetlab & 11 & 0.0005 & 0.0001 & 0.0001 & 0 & 25.75 & 1.03 \\ \hline
		Mobiact & 9 & 0.0005 & 0.0001 & 0.0005 & 0.0001 & 68.69 & 0.56 \\ \hline
		Motionsense & 11 & 0.0005 & 0.0001 & 0.0001 & 0 & 80.7 & 1.66 \\ \hline
		USC-HAD & 7 & 0.001 & 0 & 0.0005 & 0.0001 & 51.32 & 2.16 \\ \hline
		Daphnet & 9 & 0.001 & 0.00001 & 0.0005 & 0.00001 & 53.05 & 0.85 \\ \hline
		MHEALTH & 7 & 0.0001 & 0 & 0.0005 & 0.00001 & 39.2 & 1.58 \\ \hline
		PAMAP2 & 11 & 0.001 & 0 & 0.0005 & 0 & 56.88 & 2.04 \\ \hline
	\end{tabular}
\label{tab:auto_params}
\end{table}

\begin{table}[]
	\centering
	\caption{\edit{Best performing hyperparameters for SimCLR for each target dataset. }}
	\edit{
	\begin{tabular}{|c|c|c|c|c|c|c|c|}
		\hline
		Dataset & lr & wd & batch\_size & class. lr & class. L2 reg. & F1-score (mean) & F1-score (std) \\ \hline
		HHAR & 0.005 & 1e-05 & 1024 & 0.0001 & 0.0005 & 58.55 & 2.25 \\ \hline
		Myogym & 0.005 & 0.0001 & 2048 & 0.0005 & 0.0001 & 42.44 & 2.36 \\ \hline
		Wetlab & 0.005 & 0.0001 & 2048 & 0.0005 & 0.0001 & 25.93 & 2.24 \\ \hline
		Mobiact & 0.005 & 0.0 & 4096 & 0.0005 & 1e-05 & 74.89 & 1.6 \\ \hline
		Motionsense & 0.005 & 0.0 & 4096 & 0.0001 & 1e-05 & 85.6 & 2.47 \\ \hline
		USC-HAD & 0.005 & 0.0001 & 2048 & 0.0001 & 0.0005 & 53.66 & 4.12 \\ \hline
		Daphnet & 0.005 & 0.0001 & 2048 & 0.0005 & 0.0001 & 52.46 & 1.44 \\ \hline
		MHEALTH & 0.0001 & 1e-05 & 4096 & 0.0005 & 1e-05 & 50.51 & 1.16 \\ \hline
		PAMAP2 & 0.005 & 0.0001 & 2048 & 0.0005 & 0.0001 & 60.2 & 2.32 \\ \hline
	\end{tabular}
	}
	\label{tab:simclr_params}
\end{table}

\begin{table}[]
	\centering
	\caption{\edit{Best performing hyperparameters for SimSiam for each target dataset. }}
	\edit{
	\begin{tabular}{|c|c|c|c|c|c|c|c|}
		\hline
		Dataset & lr & wd & batch\_size & class. lr & class. L2 reg. & F1-score (mean) & F1-score (std) \\ \hline
		HHAR & 0.005 & 0.0 & 4096 & 0.0005 & 1e-05 & 53.22 & 2.62 \\ \hline
		Myogym & 0.001 & 0.0001 & 512 & 0.0005 & 0.0005 & 39.04 & 0.58 \\ \hline
		Wetlab & 0.0001 & 1e-05 & 4096 & 0.0005 & 1e-05 & 24.01 & 0.37 \\ \hline
		Mobiact & 0.01 & 0.0 & 2048 & 0.0005 & 0.0001 & 71.25 & 0.41 \\ \hline
		Motionsense & 0.01 & 0.0001 & 2048 & 0.0001 & 0.0001 & 83.89 & 0.65 \\ \hline
		USC-HAD & 0.005 & 0.0001 & 2048 & 0.0001 & 0.0005 & 55.46 & 1.14 \\ \hline
		Daphnet & 0.001 & 0.0001 & 1024 & 0.0005 & 0.0005 & 50.82 & 0.98 \\ \hline
		MHEALTH & 0.01 & 1e-05 & 4096 & 0.0001 & 0.0001 & 46.6 & 0.9 \\ \hline
		PAMAP2 & 0.01 & 1e-05 & 4096 & 0.0001 & 0.0001 & 59.19 & 2.07 \\ \hline
	\end{tabular}
	}
	\label{tab:simsiam_params}
\end{table}

\begin{table}[]
	\centering
	\caption{\edit{Best performing hyperparameters for BYOL for each target dataset. }}
	\edit{
	\begin{tabular}{|c|c|c|c|c|c|c|c|}
		\hline
		Dataset & lr & wd & batch\_size & class. lr & class. L2 reg. & F1-score (mean) & F1-score (std) \\ \hline
		HHAR & 0.005 & 0.0 & 4096 & 0.0005 & 1e-05 & 53.22 & 2.62 \\ \hline
		Myogym & 0.001 & 0.0001 & 512 & 0.0005 & 0.0005 & 39.04 & 0.58 \\ \hline
		Wetlab & 0.0001 & 1e-05 & 4096 & 0.0005 & 1e-05 & 24.01 & 0.37 \\ \hline
		Mobiact & 0.01 & 0.0 & 2048 & 0.0005 & 0.0001 & 71.25 & 0.41 \\ \hline
		Motionsense & 0.01 & 0.0001 & 2048 & 0.0001 & 0.0001 & 83.89 & 0.65 \\ \hline
		USC-HAD & 0.005 & 0.0001 & 2048 & 0.0001 & 0.0005 & 55.46 & 1.14 \\ \hline
		Daphnet & 0.001 & 0.0001 & 1024 & 0.0005 & 0.0005 & 50.82 & 0.98 \\ \hline
		MHEALTH & 0.01 & 1e-05 & 4096 & 0.0001 & 0.0001 & 46.6 & 0.9 \\ \hline
		PAMAP2 & 0.01 & 1e-05 & 4096 & 0.0001 & 0.0001 & 59.19 & 2.07 \\ \hline
	\end{tabular}
	}
	\label{tab:byol_params}
\end{table}

\begin{table}[]
	\centering
	\caption{Best performing hyperparameters for DeepConvLSTM for each target dataset.}
	\begin{tabular}{|c|c|c|c|c|}
		\hline
		Dataset & lr & wd & F1-score (mean) & F1-score (std) \\ \hline
		HHAR & 0.0005 & 0 & 54.39 & 2.28 \\ \hline
		Myogym & 0.0005 & 0.00001 & 39.9 & 1.05 \\ \hline
		Wetlab & 0.001 & 0 & 31 & 0.68 \\ \hline
		Mobiact & 0.001 & 0.00001 & 82.21 & 0.69 \\ \hline
		Motionsense & 0.0001 & 0.0001 & 84.56 & 0.85 \\ \hline
		USC-HAD & 0.0005 & 0.0001 & 53.64 & 0.51 \\ \hline
		Daphnet & 0.0001 & 0.00001 & 53.68 & 2.58 \\ \hline
		MHEALTH & 0.0005 & 0 & 45.91 & 0.89 \\ \hline
		PAMAP2 & 0.001 & 0 & 51.22 & 1.91 \\ \hline
	\end{tabular}
\label{tab:deepconv_params}
\end{table}

\begin{table}[]
	\centering
	\caption{Best performing hyperparameters overall, across all target datasets.}
	\begin{tabular}{|c|c|}
		\hline
		Approach & Best overall parameters \\ \hline
		Multi-task & lr=0.0003, l2 reg.=0.0005, class. lr=0.0003, class. l2 reg.=0 \\ \hline
		Masked recons. & head=8, layers=6, mask\%=10, warmup=80000, class. lr=0.0005, class. l2 reg.=0.0001 \\ \hline
		CPC & lr=0.0001, l2 reg.=0.00001, kernel\_size=3, k=64, bs=256, class. lr=0.0005, class. l2 reg.=0 \\ \hline
		Autoencoder & lr=0.001, wd=0.00001, kernel\_size=9, class. lr=0.0005, class. l2 reg.=0.00001 \\ \hline
		\edit{SimCLR} & \edit{lr=0.005, wd=0.0001, batch\_size=2048, class. lr=0.0001, class. l2 reg.=0.0005} \\ \hline
		\edit{SimSiam} & \edit{lr=0.0005, wd=0.0001, batch\_size=256, class. lr=0.0005, class. l2 reg.=0.0005} \\ \hline
		\edit{BYOL} & \edit{lr=0.01, wd=0.00001, batch\_size=4096, class. lr=0.0005, class. l2 reg.=0} \\ \hline
	\end{tabular}
\label{tab:overall_params}
\end{table}

\begin{figure}[h]
	\centering
	\includegraphics[width=0.42\textwidth]{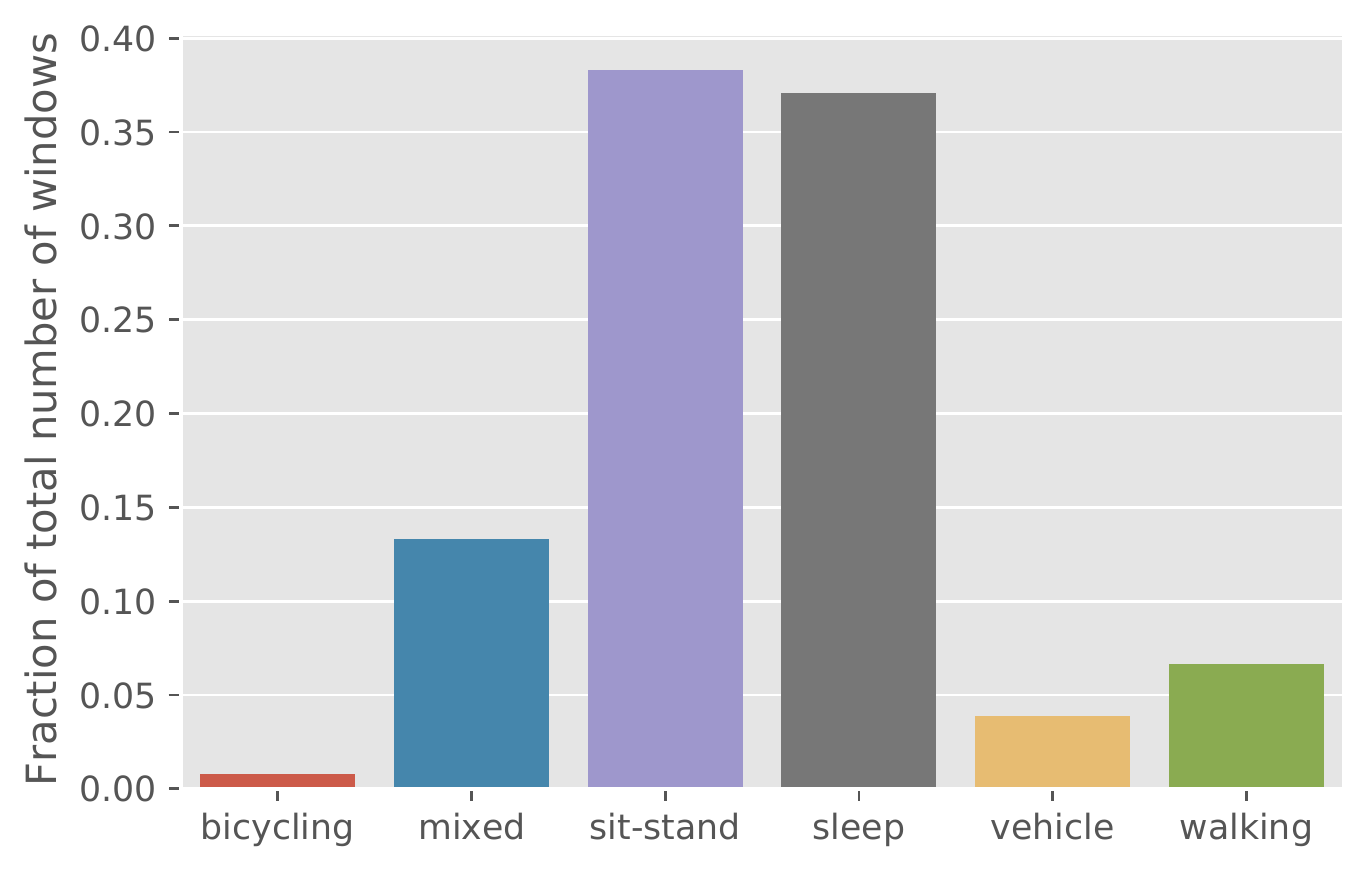}
	\caption{
		The class composition of the Capture-24 train dataset, demonstrating that it is imbalanced. 
		Sleep and sit-stand comprise around 75\% of the total dataset, whereas bicycling only consists of 0.78\% of the train data.
	}
	\label{fig:capture_24_composition}
\end{figure}

\begin{figure}[h]
	\centering
	\includegraphics[width=\textwidth]{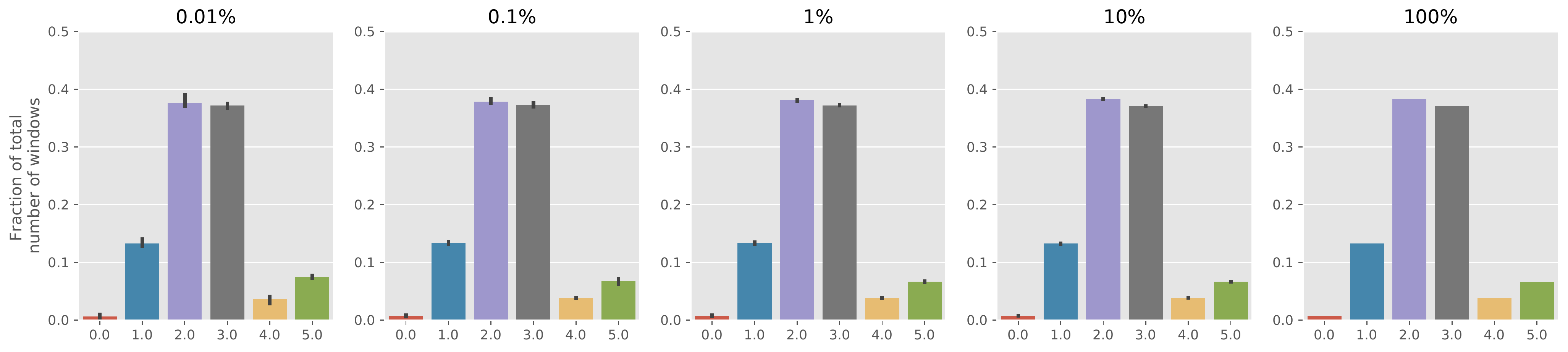}
	\caption{
		\edit{
			When the training windows are sub-sampled based on a percentage, the class distribution generally follows the composition of the full train dataset (see Fig. \ref{fig:capture_24_composition}) albeit the number of samples in each class are different.
			We show the distribution for three random seeds, which also help reduce the impact of only picking the majority class.
		}
	}
	\label{fig:perc_distribution}
\end{figure}

\begin{figure*}[t]
	\centering
	\includegraphics[width=0.8\textwidth]{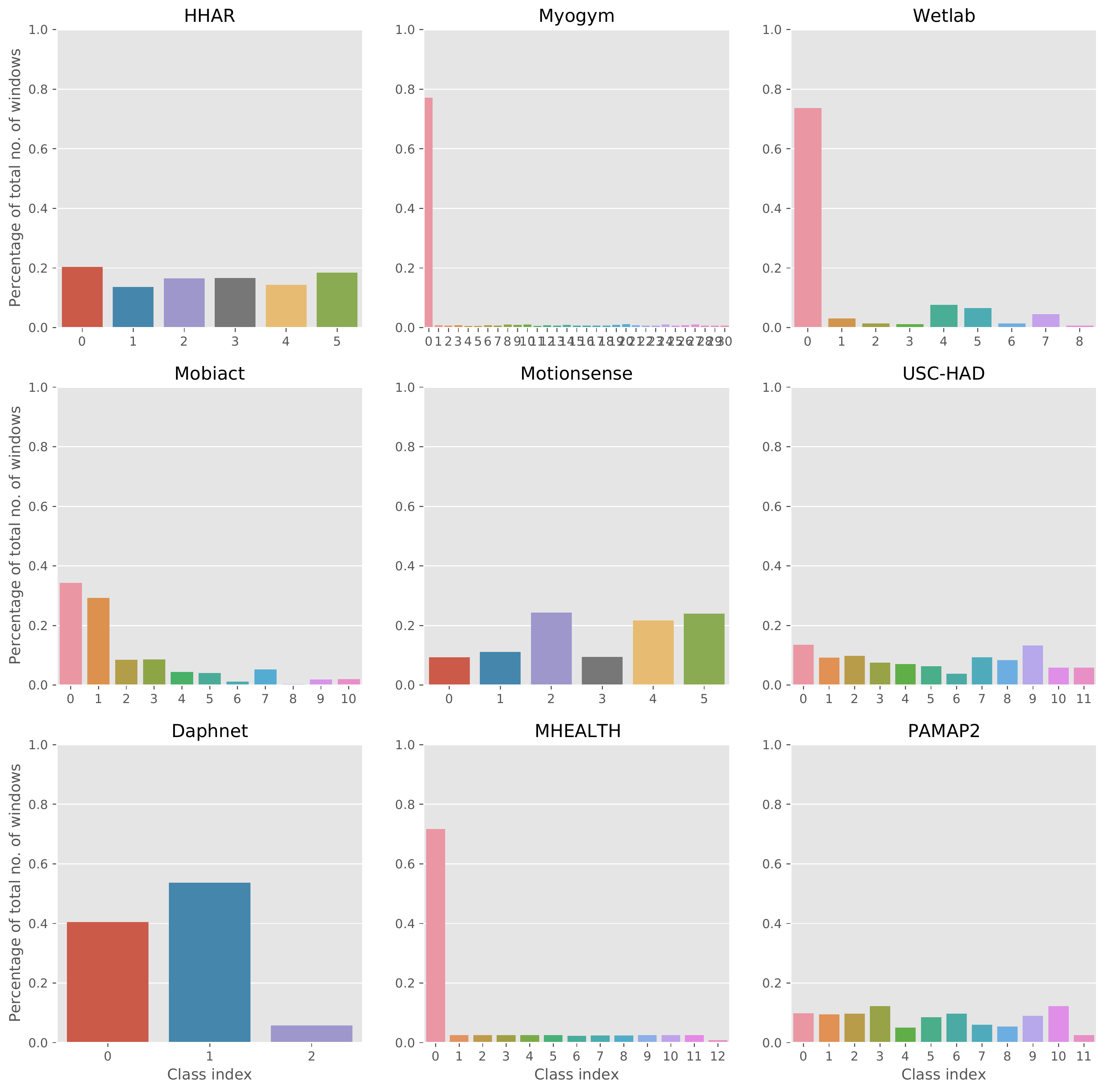}
	\caption{
		The class composition of all target datasets.
	}
	\label{fig:target_composition}
\end{figure*}

\subsection{Transformation Functions used for SimCLR, SimSiam, and BYOL}
For contrastive learning with SimCLR, SimSiam, and BYOL, we utilized eight time-series transformation functions proposed in \cite{um2017data}. 
The parameters for these transformations were obtained from \cite{tang2020exploring}, available on Github: \url{https://github.com/iantangc/ContrastiveLearningHAR/blob/main/transformations.py}

\begin{enumerate}
	\item Jitter: random Gaussian noise is added to each channel with a mean of zero and standard deviation of 0.05.
	\item Scaling: each channel is multiplied by a random value from a normal distribution of mean=1.0 and standard deviation=0.1.
	\item Rotation: a 3D rotation matrix is applied based on a randomly chosen axis and randomly chosen rotation angle (both are drawn from the uniform distribution).
	\item Permutation: the three channels are randomly permuted.
	\item Scrambling: the window of data is broken into 4 sections, each of which are permuted randomly, and then recombined.
	\item Time warping: four fixed points are used to generate a cubic spline. The window is then streched and warped across time based on the cubic spline, which controls the deviation of the speed of time relative to normal. 
	\item Negation: each channel is multipled by -1.
	\item Reversing: the entire window is reversed in time.
\end{enumerate}

\subsection{Source Dataset Imbalance Setup}
\label{app:source_data_imb}
\edit{
	The imbalance factor $\rho$ is given by:
	$$ \rho = \frac{\text{num\_windows of the rarest class}}{\text{num\_windows of the most frequent class}} \leq 1 $$
	For example, if we have an imbalance ratio $\rho=0.01$, the rarest class will have 200 windows, whereas the most frequent class has 20,000 windows, and the remaining classes are populated according to the exponential. 
	Therefore, the total number of windows is dependent on the class imbalance ratio. 
}

\edit{
	For every $\rho$, the balanced subset divides the total number of windows for the corresponding imbalanced subset by the number of classes in Capture-24, which is $6$. 
	For example, given a $\rho$ of $0.001$, the total number of windows is $\sim$27k in the imbalanced subset, and thus, each class of the balanced subset contains $\sim$4.5k windows (resulting in a total of $\sim$27k windows).  
}
\newpage

\begin{figure}[h]
	\centering
	\includegraphics[height=0.9\textheight]{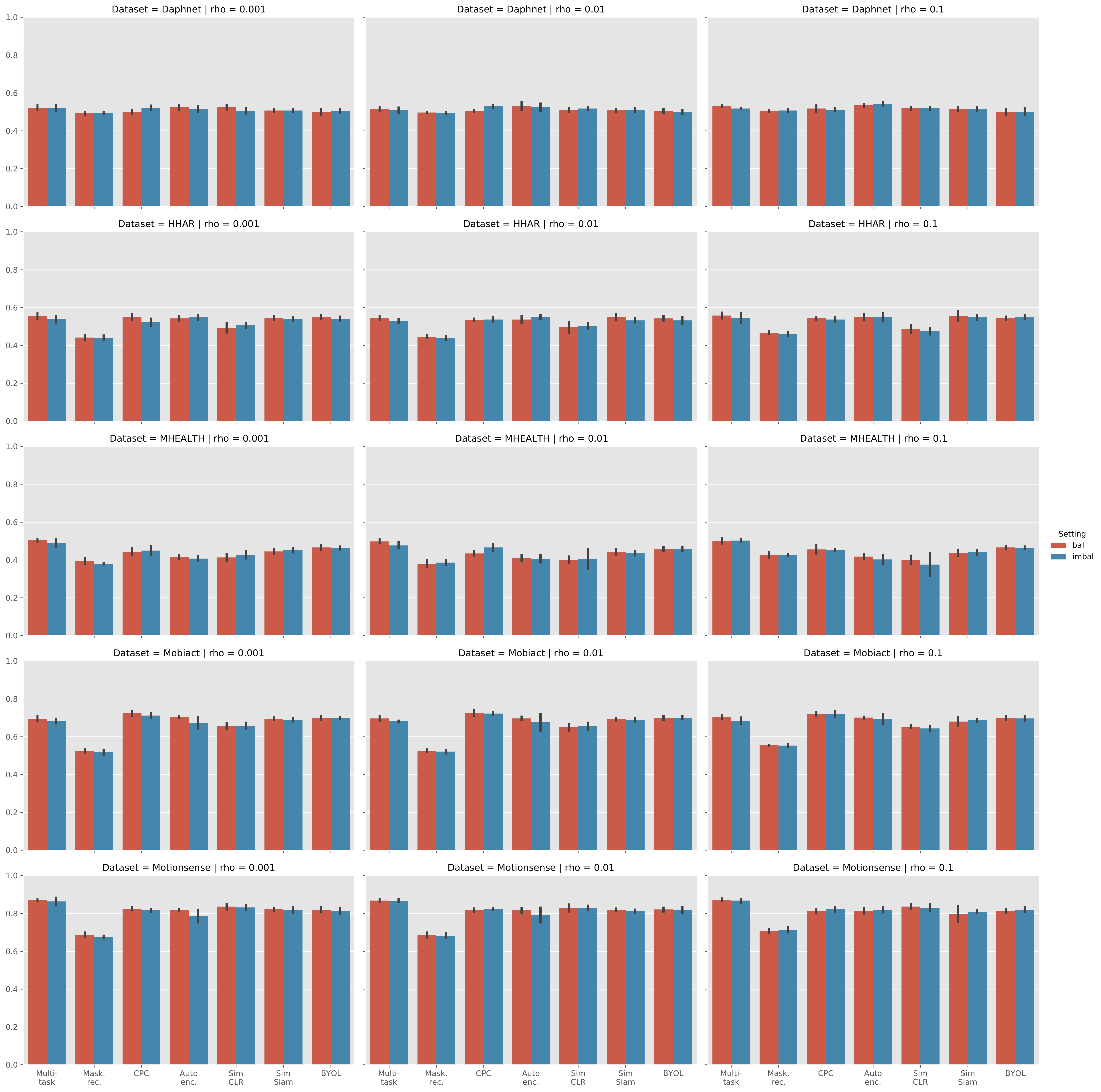}
	\caption{
		Impact of the source dataset imbalance on the activity recognition performance - Part 1.
	}
	\label{fig:first_source_imbalance}
\end{figure}

\begin{figure}[h]
	\centering
	\includegraphics[height=0.7\textheight]{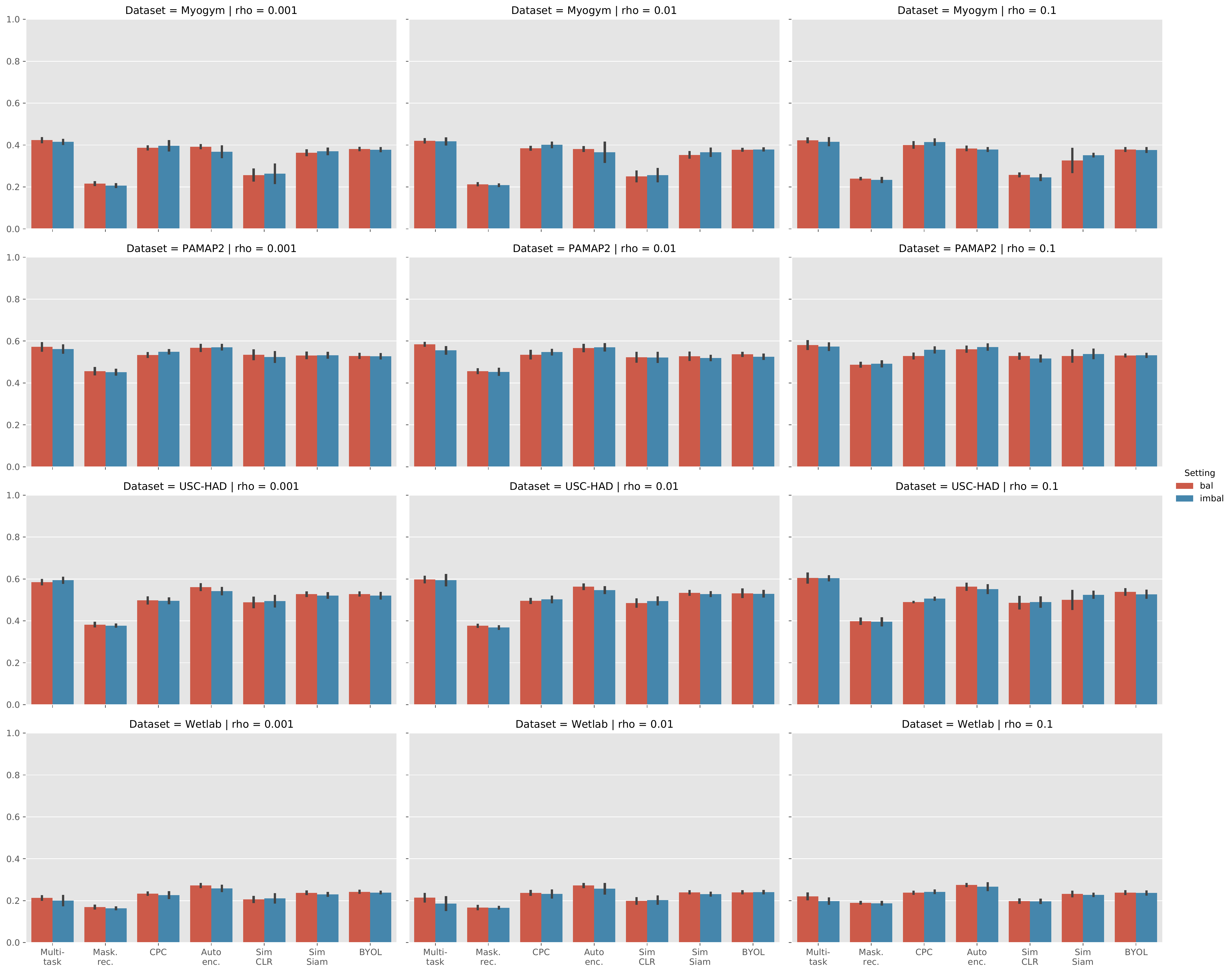}
	\caption{
		Impact of the source dataset imbalance on the activity recognition performance - Part 2.
	}
	\label{fig:second_source_imbalance}
\end{figure}

\begin{figure}[t]
	\centering
	\includegraphics[width=\textwidth]{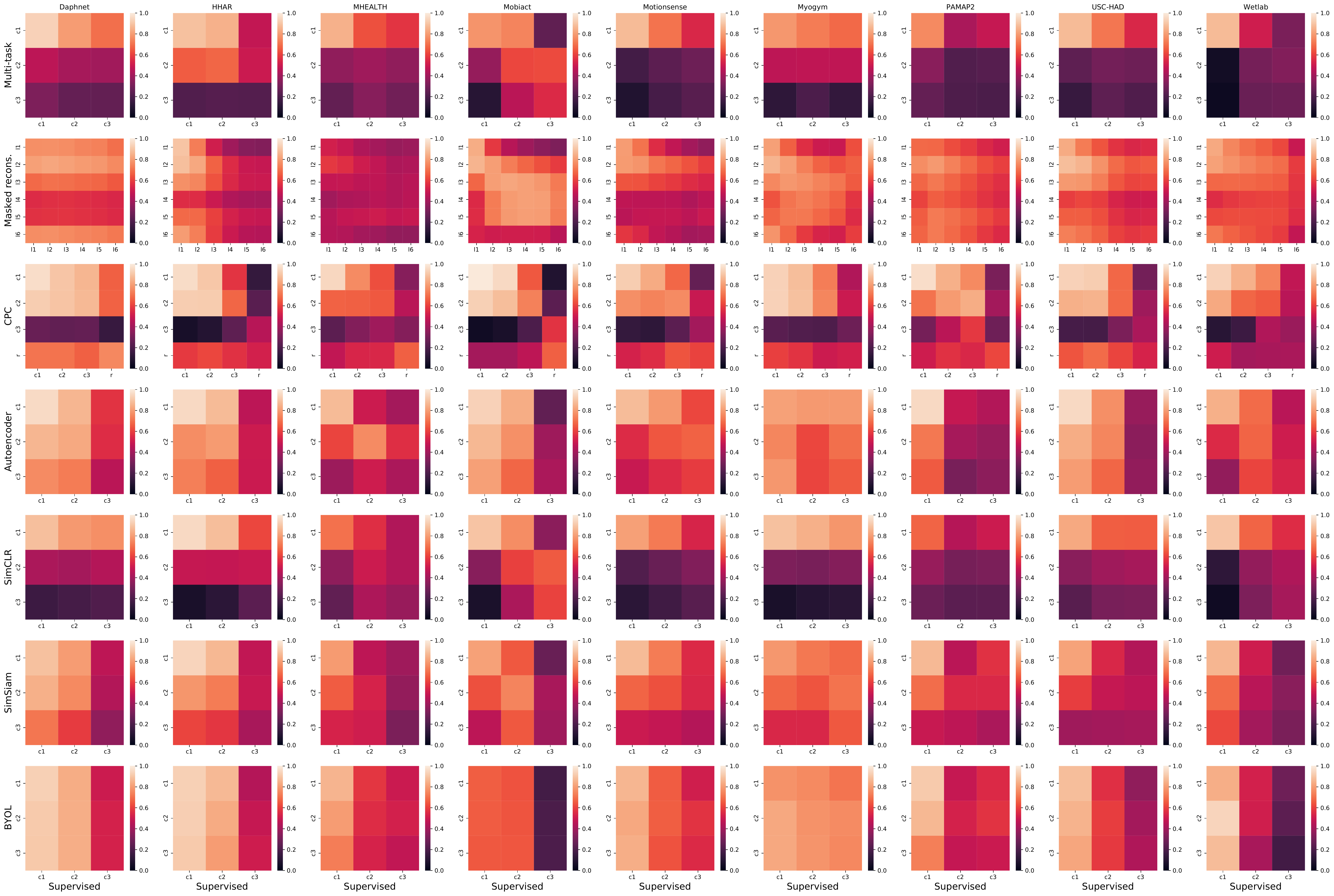}
	\caption{
		Similarity of the learned representations to supervised learning.
	}
	\label{fig:all_similarity}
\end{figure}

\begin{figure}[t]
	\centering
	\includegraphics[width=\textwidth]{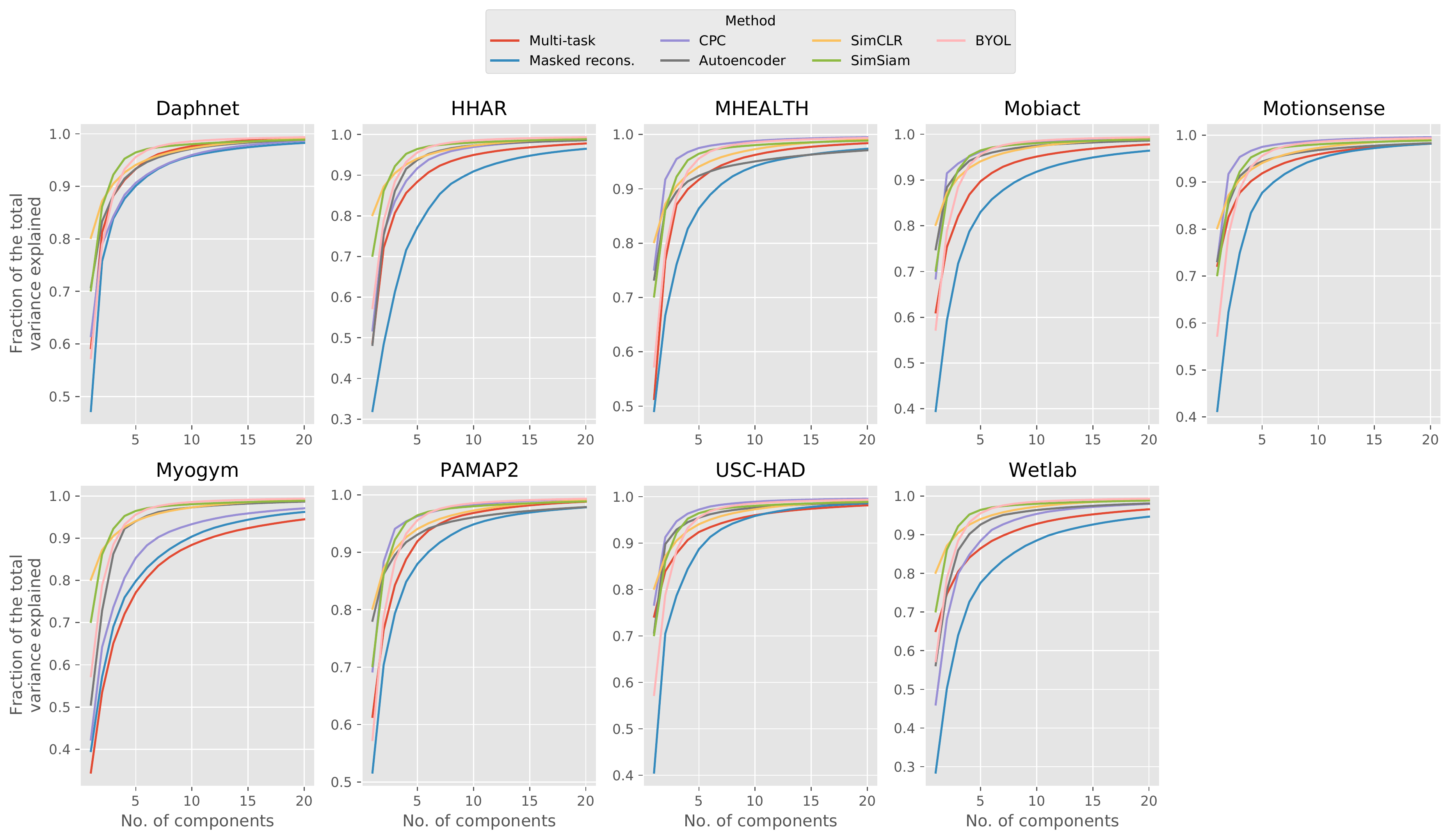}
	\caption{
		Implicit dimensionality of the self-supervised approaches for individual target datasets.
	}
	\label{fig:all_imp_dim}
\end{figure}

\begin{figure}[t]
	\centering
	\includegraphics[width=\textwidth]{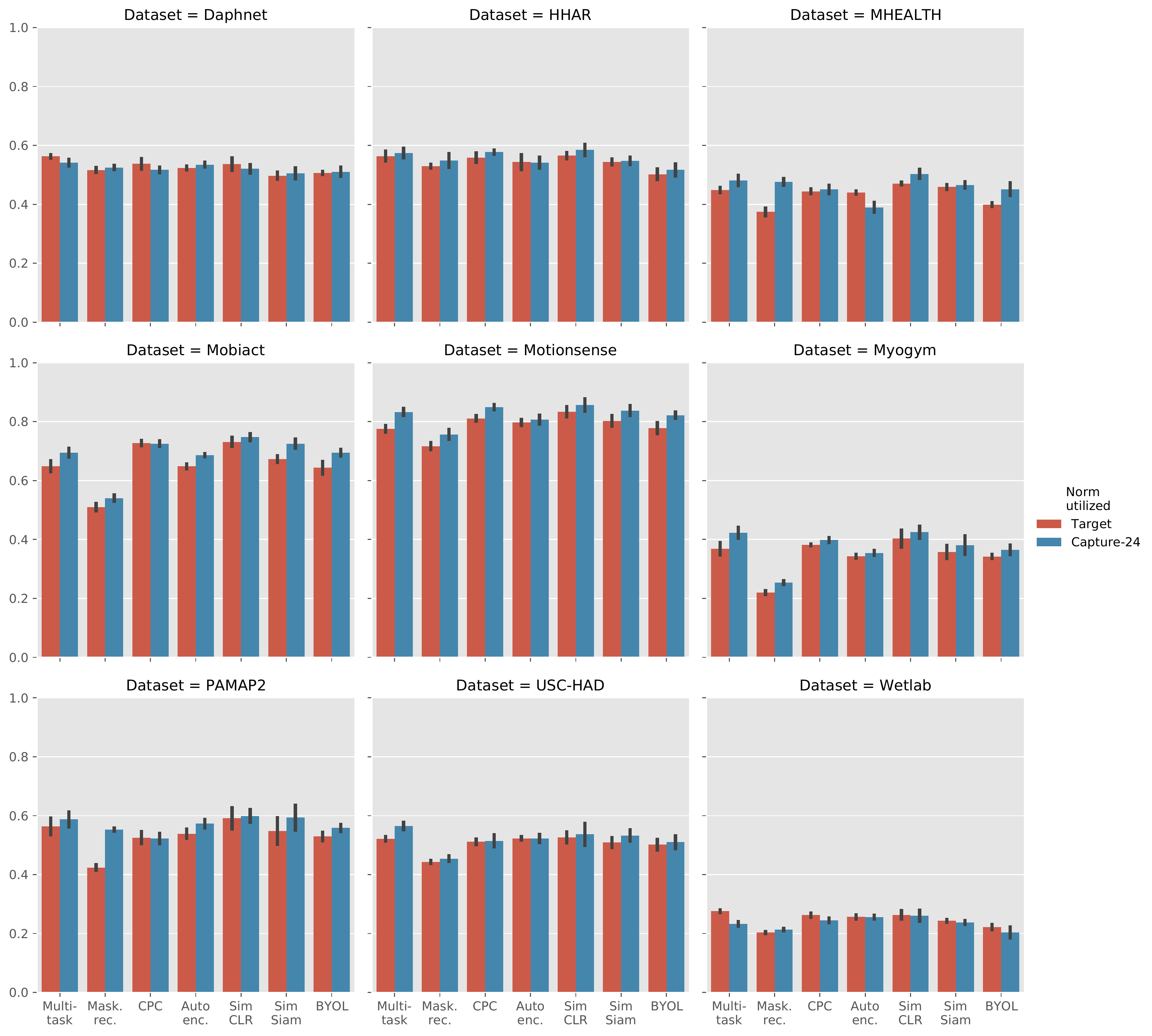}
	\caption{
		Activity recognition performance when utilizing either the source dataset means and variances, or the target dataset statistics.
	}
	\label{fig:all_cap_norm}
\end{figure}

\end{document}